\documentclass[11pt,onecolumn,floatfix,altaffilletter,superscriptaddress,tightenlines,showpacs,showkeys,preprintnumbers,nofootinbib]{revtex4-2}
\pdfoutput=1
\usepackage[colorlinks=true,citecolor=blue,linkcolor=blue,breaklinks=true,linktoc=none]{hyperref}
\usepackage{amsmath,amssymb}
\usepackage{epsfig} 
\usepackage{graphicx}
\usepackage{url}
\usepackage{color}
\usepackage{multirow}
\usepackage{placeins}
\usepackage[dvipsnames]{xcolor}
\usepackage{braket}
\usepackage{float}
\usepackage{slashed}
\usepackage{mathtools}

\hypersetup{
  colorlinks=true,
  linkcolor=red,
  filecolor=magenta,   
  urlcolor=blue,
  citecolor=blue,
}

\hypersetup{colorlinks,linkcolor={blue},citecolor={teal},urlcolor={violet}}

\allowdisplaybreaks

\bibpunct{[}{]}{,}{n}{}{,}

\def\beq{\begin{equation}}
\def\eeq{\end{equation}}
\def\bea{\begin{eqnarray}}
\def\eea{\end{eqnarray}}
\makeatletter
\@addtoreset{equation}{section}

\newcommand{\INFN}{INFN - Sezione di Napoli, Complesso Universitario Monte S. Angelo, I-80126 Napoli, Italy}

\newcommand{\SSM}{Scuola Superiore Meridionale, Università degli studi di Napoli ``Federico II'', Largo San Marcellino 10, 80138 Napoli, Italy}

\begin{document}

\title{Probing Leptogenesis at LISA: A Fisher analysis} 
 \author{Rome Samanta}
  \email{samanta@na.infn.it}
   \email{romesamanta@gmail.com}
  \affiliation{\SSM}
  \affiliation{\INFN}

\begin{abstract}
In \textbf{JCAP 11 (2024) 051}, we discussed how different regimes (flavoured) of leptogenesis can be probed through a ``tomographic'' approach using primordial gravitational waves. By examining the theory's parameter space, we identified regions where right-handed neutrino mass-dependent non-standard cosmological expansion leaves characteristic imprints on propagating gravitational waves. Our analysis focused on inflationary blue-tilted gravitational waves, modeled by a power-law tensor power spectrum with a constant spectral index. The resulting double-peak spectrum---where peak and dip frequencies are sensitive to leptogenesis parameters---provided marked signatures of different leptogenesis regimes.  In this follow-up article, we conduct a statistical analysis of two-flavour leptogenesis signals, particularly those producing a peak (more generally, a broken power-law signal) within the LISA frequency band. Using a Fisher matrix analysis, we delineate the regions of parameter space that LISA can probe with minimal uncertainty, accounting for galactic and extragalactic foregrounds along with LISA's instrumental noise.

\end{abstract}
\maketitle
\tableofcontents

\section{Introduction}
{\color{black}Inflationary Blue-tilted gravitational waves (BGWs) that might arise in scenarios such as Non-Slow-Roll Inflation models, Modified Gravity (Galileon-type interaction term and the Gauss-Bonnet coupling) \cite{bgw1,bgw2,bgw3,bgw4,bgw5,bgw6,bgw7,bgw8,bgwnew,Tahara:2020fmn,Piao:2004tq,Caldwell:2017chz,Datta:2023xpr}, have recently gained attention following the acclaimed discovery of stochastic gravitational waves by the Pulsar Timing Array (PTA) ~\cite{ng1,ng2,ng3,ng4,ng5,EPTA:2023xxk,bn1n,bn2n,bn3n,bn4n,bn5n,bn6n}.} A fit to the data suggests that BGWs, modeled as a power-law spectrum, require an extremely large spectral index $n_T\sim 1.8$ and a low reheating temperature after inflation {\color{black}($T_R\lesssim 10$ GeV)} \cite{ng5}. {\color{black} The latter arises because the energy density of BGWs} must not violate the Big Bang Nucleosynthesis (BBN) constraints on the effective number of neutrino species~\cite{Peimbert:2016bdg}. However, a post-inflationary matter-dominated era can generate entropy and suppress BGWs to such an extent that they not only evade BBN constraints but also span multiple decades in frequency, exhibiting characteristic spectral features determined by the properties of the intermediate matter domination \cite{bn3n,t1,t2,t3,t4,t5,t6}. This presents an intriguing opportunity for beyond Standard Model (BSM) physics, enabling a tomographic approach to probe the early universe \cite{Datta:2022tab,Datta:2023vbs,Chianese:2024nyw}. 

In Ref.~\cite{Chianese:2024nyw}, it has been discussed that realistic seesaw models introduce an early matter-dominated phase ({\color{black}see, e.g., Ref.s \cite{mat1,mat2} for overarching discussions on early matter domination and GWs}) driven by the oscillation of a scalar field, which gives mass to right-handed neutrinos (RHNs). Within certain parameter spaces, the duration and the end of this matter domination correlate with RHN masses ($M_{Ni}$), which typically set the scale of thermal leptogenesis: {\color{black} $T_{\rm lepto}$ \cite{lep2,lep3,lep4,lep5,lep6,lep7,lep8,lep9,lep11,lep12}.} As BGWs propagate through this RHN mass-dependent matter epoch, they encode information about the leptogenesis scale. This mechanism offers a novel way to probe high-scale leptogenesis, which is otherwise inaccessible in collider experiments, unlike the low-scale scenarios \cite{lep3,lep9,lep11}.  

Furthermore, since high-scale leptogenesis operates in different flavour regimes \cite{fllep1,fllep2,fllep3}—each characterized by specific RHN mass ranges—distinct signals arise~\cite{Chianese:2024nyw}. {\color{black}Due to matter domination and the subsequent entropy production, these models lead to the suppression of the BGW spectrum with a detectable double-peak signature:} the first peak appears in the $\mu$Hz range for a three-flavour regime ($M_{Ni}\lesssim 10^9\,\text{GeV}$), while for a two-flavour regime ($10^9\,\text{GeV}\lesssim M_{Ni}\lesssim 10^{12}\,\text{GeV}$), { \color{black}it shifts to the mHz range, within LISA’s sensitivity \cite{Baker:2019nia,LISACosmologyWorkingGroup:2022jok,LISA:2024hlh}. This article presents a numerical follow-up to Ref.~\cite{Chianese:2024nyw}, focusing on the two-flavour leptogenesis scenario and the associated BGW signal, which could potentially be detected by LISA with a peak amplitude as large as  $\Omega_{\rm GW}\sim 10^{-8}$.}

The LISA mission, adopted by the European Space Agency (ESA), is scheduled for launch in the 2030s. Despite LISA's immense physics potential, detecting a primordial stochastic gravitational wave background remains highly challenging. The data stream will be contaminated by instrumental noise, unresolved astrophysical sources (collectively treated as stochastic foregrounds), and numerous transient events overlapping in time.

In this article, we shall assume that the LISA data will contain instrumental noise plus galactic and extragalactic foreground along with the primordial signal provided by the leptogenesis model.  While the galactic foreground primarily originates from white dwarf binaries within our galaxy  \cite{Nissanke:2012eh,Evans:1987qa,Cornish:2017vip},  {\color{black}neutron stars, stellar-origin black hole binaries emitting gravitational waves during their inspiral phase, and white dwarf binaries (recent studies suggest that the contribution from these white dwarfs could dominate in the LISA frequency band \cite{Staelens:2023xjn,Hofman:2024xar}. This result, however, has not been included in our analysis), may source extragalactic foreground \cite{Regimbau:2011rp,Babak:2023lro,Lehoucq:2023zlt}.} Given all these components, we compute the uncertainty in the parameters describing the primordial signal with a $8\times 8$ Fisher matrix (4 primordial signal parameters, 2 instrumental noise parameters, and 2 astrophysical foreground parameters) analysis adopting a single channel (TD1X) noise spectra \cite{Armano:2018kix,LISAsc}. 

We find that even when maximizing the primordial signal with respect to the tensor-to-scalar ratio and assuming precisely known noise and foregrounds, reducing the uncertainty in the primordial signal parameters requires a reasonably large tensor blue-tilt of $n_T \sim 0.45$. This value increases to $\sim 0.5$ when uncertainties in all parameters are considered.\footnote{The precision of these values can be further refined through a multichannel approach (e.g., XYZ and AET) or a Bayesian analysis using Markov Chain Monte Carlo
(MCMC) simulations \cite{Flauger:2020qyi}.} In this model, such values of the spectral index imply that the primordial signal strength must exceed instrumental noise sensitivity within a certain frequency range.  Beyond serving as a testable framework for generating a leptogenesis model-imprinted GW signal, this scenario thus also underscores the importance of uncertainty analysis in primordial GW predictions in BSM models. This is particularly relevant as theoretical studies, including Ref.~\cite{Chianese:2024nyw}, often assume that a model can be accurately probed with weak signals—i.e., once the associated GW spectrum intersects the Power-Law integrated Sensitivity (PLS) curves \cite{Thrane:2013oya,Schmitz:2020syl}. However, this assumption may not always hold. In particular, computing only Signal-to-Noise-Ratio (SNR) and showing it to be more than 1, or even 10, may not be the showdown to assess the detectability of GW signals for BSM models.

\begin{figure}[t!]
\centering
\includegraphics[scale=.48]{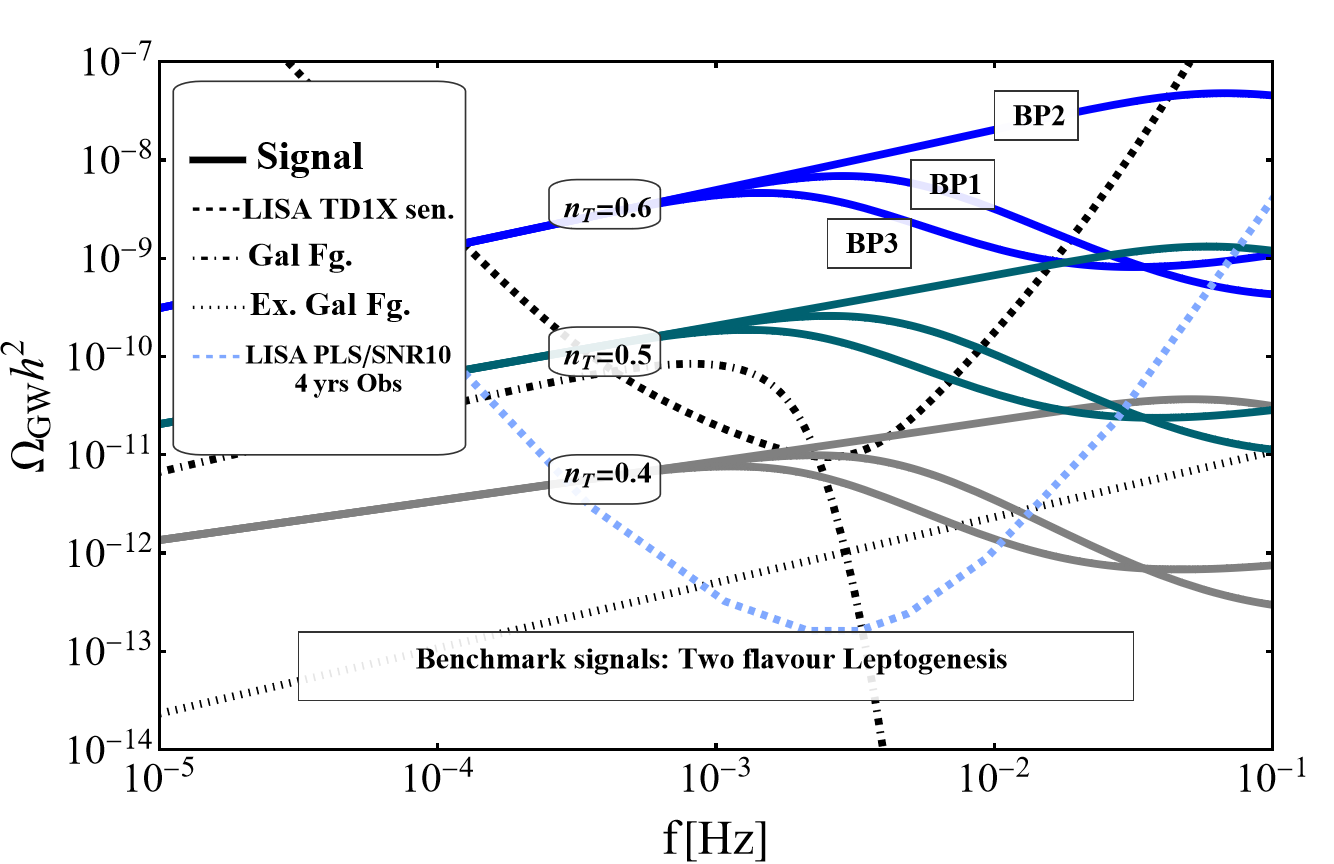}
\caption{\color{black}Sets of BGW spectra for a two-flavour leptogenesis scenario: $10^9\,\text{GeV} \lesssim M_N \sim T_{\rm lepto} \lesssim 10^{12}\,\text{GeV}$. The sets with blue, dark orange, and purple curves correspond to tensor spectral indices $n_T = 0.6$, $0.5$, and $0.4$, respectively. For all the displayed spectra, Eq.\eqref{omgGW} has been used with the tensor power spectrum parameterized as a power-law in Eq.\eqref{Pt} and evaluated at $r=0.06$.  Each set contains three GW spectra, corresponding to the leptogenesis model benchmark points (BPs) listed in Table~\ref{t1}. The figure also includes the noise curve (black dashed), PLS curve (blue dashed), galactic foregrounds (dot-dashed), and extragalactic foregrounds (dotted).}
\label{fig:fig} 
\end{figure}
For readers interested only in the introduction of this article, in Fig.~\ref{fig:fig}, we present sets of representative GW spectra (with model parameter values discussed later) alongside the noise (black dashed), {\color{black} PLS curve (blue dashed) with SNR = 10 and observation time = 4 years, and foregrounds (dot-dashed, dotted).} {\color{black}As discussed in the next sections, the spectra shown in blue are associated with lower parameter uncertainties, indicating a reconstructible leptogenesis-BGW scenario within the LISA frequency band. In contrast, as the signal strength decreases (from dark orange to purple curves), the model becomes increasingly susceptible to foreground contamination and instrumental noise, making them weakly reconstructible scenarios. We also refer the reader to Appendix~\ref{appn1} for an explanatory overview of various aspects of the model.
}

The rest of the paper is organized as follows: In Sec.~\ref{s2}, we briefly outline the model. In Sec.~\ref{s3}, we discuss the properties of BGWs and their connection to leptogenesis. In Sec.~\ref{s4}, we present the Fisher forecast analysis. Finally, we conclude in Sec.~\ref{s5}.

\section{The model parameter space responding to primordial gravitational waves}\label{s2}

\textbf{General Setup:} As outlined in Ref.\cite{Chianese:2024nyw}, this model is based on $U(1)_{B-L}$ symmetry embeddable in Grand Unified Theories (GUT) \cite{bml1,bml2,bml3,bml4,bml5}. The Lagrangian governing the leptogenesis process and the dynamics of the scalar field is given by  

\bea
-\Delta \mathcal{L}\subset f_{D i\alpha} \overline{N_{Ri} }\tilde{H}^\dagger L_{\alpha}+\frac{1}{2}f_{Ni}\overline{N_{Ri}}\Phi N_{Ri}^C+\frac{1}{4}\lambda_{H\Phi}|H|^2|\Phi|^2+V(\Phi,T)\,,\label{lag}
\eea

where \( f_D \) is the neutrino Dirac Yukawa coupling, \( H \) (\( L \)) represents the Standard Model (SM) Higgs (lepton doublet), \( N_R \) is the  RHN field, and \( \Phi \) is a scalar field with a vacuum expectation value \( v_\Phi \). The finite temperature potential \( V(\Phi, T) \) dictates the phase transition dynamics of \( \Phi \). The fields \( N_R \) and \( L \) carry a \( B - L \) charge of \(-1\), whereas \( \Phi \) and \( H \) have charges of \( 2 \) and \( 0 \), respectively.  

A key aspect of this model is the Higgs portal coupling \( \lambda_{H\Phi} \), which is given by  

\begin{equation}
\lambda_{H\Phi} \simeq \lambda_{H\Phi}^{\text{tree}} + \lambda_{H\Phi}^{\text{1-loop}},
\end{equation}

where \( \lambda_{H\Phi}^{\text{1-loop}} \) arises from the first two terms in {\color{black}Eq.\eqref{lag}} and induces the decay \( \Phi \to hh \) (see Fig.2 of Ref.\cite{Chianese:2024nyw}). We assume \( \lambda_{H\Phi}^{\text{tree}} \lesssim \lambda_{H\Phi}^{\text{1-loop}} \) across all relevant physical scales.{ \color{black}This assumption is central and links neutrino parameters with   \(\Phi \)'s decay width $  \Gamma_\Phi^{hh}$ given in Eq.\eqref{newgamma}.   }

As the temperature decreases, the scalar field transitions from \( \Phi = 0 \) to its vacuum expectation value \( \Phi = v_\Phi \). Consequently, the RHNs acquire mass 

\begin{equation}
 \color{black}    M_i = f_{N_i} v_\Phi.
\end{equation}

 After the electroweak phase transition (\( T \lesssim T_{\text{EW}} \)), the Higgs field stabilizes at its vacuum expectation value \( H = v_h \), and the first two terms in the Lagrangian generate light active neutrino masses via the standard type-I seesaw mechanism,  

\begin{equation}
m_{\nu_i} \sim \frac{f_{D_i}^2 v_h^2}{M_i}.\label{nueq}
\end{equation}

At high temperatures (\( T \gg T_{\text{EW}} \)), heavy RHNs decay CP-asymmetrically into lepton doublets and Higgs bosons, generating a lepton asymmetry around \( T \sim M_i \)—the scale of thermal leptogenesis. For simplicity, we assume a single RHN mass scale for the rest of the discussion, i.e., \( M_i \equiv M_N = f_N v_\Phi \), without focusing on the mass hierarchy among \( M_i \).  {\color{black}Note, however, that the scenario can be trivially generalized to include hierarchical RHN masses. In such cases, leptogenesis is primarily driven by the decay of the lightest RHN, since any asymmetry generated by the heavier states is subsequently washed out via inverse decays involving the lightest one \cite{lep4}. On the other hand, the lifetime of the scalar field $\Phi$ is determined by the largest Yukawa coupling ($f_N$), and thus by the mass of the heaviest RHN. As a result, the mass hierarchy between the lightest and heaviest RHNs effectively links the leptogenesis scale to the decay timescale of $\Phi$. In the context of parameter reconstruction, it would therefore be essential to treat the mass hierarchy as an additional free parameter. Another interesting aspect of this model is the possibility of generating an additional period of matter domination by making one of the RHNs long-lived. Since only two RHNs are sufficient to successfully seed leptogenesis and explain neutrino oscillation data \cite{Esteban:2024eli}, one could, for example, take the Yukawa coupling $f_D$ (in Eq.~\eqref{lag}) corresponding to one RHN (e.g., the third) to be small, thereby delaying its decay \cite{Asaka:2020wcr,Borah:2022byb}. This would naturally lead to a secondary phase of matter domination. We do not adopt such an assumption in the present analysis. This scenario, however, must be approached with caution. The late decay of the long-lived RHN would inject additional entropy, causing further dilution of the baryon asymmetry—effectively resulting in a double dilution: first due to the scalar field, and second due to the RHN itself. }

The finite-temperature potential responsible for restoring symmetry at higher temperatures is given by \cite{Linde:1978px,Kibble:1980mv,Quiros:1999jp,Caprini:2015zlo,Hindmarsh:2020hop}:  

\begin{equation}
V(\Phi,T) = D(T^2 - T_0^2)\Phi^2 - E T \Phi^3 + \frac{\lambda}{4} \Phi^4, \label{tmpdp}
\end{equation}

where the coefficients are defined as  

\begin{equation}
D = \frac{3 g^{\prime 2} + 4\lambda}{24}, \quad
E = \frac{3 g^{\prime 3} + g^\prime \lambda + 3 \lambda^{3/2}}{24 \pi}, \quad
T_0 = \frac{\sqrt{12\lambda} v_{\Phi}}{\sqrt{3 g^{\prime 2} + 4\lambda}}.
\end{equation}

Here, \( g^\prime \) is the gauge coupling, and the vacuum expectation value \( v_\Phi = \mu / \sqrt{\lambda} \) is determined from the zero-temperature potential  

\begin{equation}
V(\Phi,0) = -\frac{\mu^2}{2} \Phi^2 + \frac{\lambda}{4} \Phi^4.
\end{equation}

The second  last term in Eq.~\eqref{tmpdp} creates a potential barrier, leading to a second minimum at \( \Phi \neq 0 \). At \( T = T_c \), this barrier causes the minimum at \( \Phi = 0 \) to degenerate with the second minimum. Around \( T_0 \) (\( T_0 \lesssim T_c \)), the potential barrier disappears, and the minimum at \( \Phi = 0 \) becomes a maximum~\cite{Quiros:1999jp}. The critical temperature \( T_c \) and the corresponding field value \( \Phi_c = \Phi (T_c) \) are given by~\cite{Quiros:1999jp,Megevand:2016lpr}:  

\begin{equation}
T_c = T_0 \frac{\sqrt{\lambda D}}{\sqrt{\lambda D - E^2}}, \quad
\Phi_c = \sqrt{\frac{4 D}{\lambda} (T_c^2 - T_0^2)}.
\end{equation}

The strength of the phase transition is characterized by the order parameter \( \Phi_c / T_c \) \cite{Quiros:1999jp}. When \( \Phi_c / T_c \ll 1 \), the potential barrier disappears rapidly (\( T_c \simeq T_0 \)), and the transition is effectively described by the rolling of the field \( \Phi \) from \( \Phi = 0 \) to \( \Phi = v_\Phi \). We impose parameter constraints ensuring \( \Phi_c / T_c \ll 1 \), which is satisfied for\footnote{The coupling hierarchy \( \lambda \simeq g^{\prime (n=3)} \) serves as a useful rule of thumb. In this model, the most sensitive parameter is the coupling \( f_N \). For larger values of \( n \), the potential barrier height increases, making it challenging to achieve a smooth rolling of the field. Conversely, for smaller values of \( n \), constraints associated with gauge couplings become more restrictive. As a result, \( n=3 \) emerges as the optimal choice for the model. While one may consider a region around \( n=3 \), the qualitative features of the model would remain largely unchanged. See, e.g., Ref.~\cite{Chianese:2024gee} for a discussion.
} \( \lambda \simeq g^{\prime 3} \) and \( g^\prime \lesssim 10^{-2} \). This choice of parameters leads to a critical temperature  

\begin{equation}
T_c \simeq T_0 = 2\sqrt{g^\prime} v_\Phi.
\end{equation}

Additionally, two important constraints must be satisfied:
\begin{itemize}
    \item \textbf{No second period of inflation:} We work in a parameter space such that the energy density of the scalar field at \( T_c \)  remains subdominant to the radiation energy density,  
    \begin{equation}
    \rho_\Phi (T_c) \sim \frac{\lambda v_\Phi^4}{4} < \rho_R (T_c).
    \end{equation}
    
  {\color{black}  \item \textbf{Oscillation condition:} The mass of \( \Phi \) must exceed the Hubble parameter at \( T_c \),  
    \begin{equation}
    m_\Phi = \sqrt{2\lambda} v_\Phi \gtrsim \mathcal{H}(T_c),
    \end{equation}
    so that the field oscillates at the bottom of the potential to provide matter domination.}
\end{itemize}

These constraints simplify to  

\begin{equation}
\rho_\Phi (T_c) < 4.5 \times 10^{-6} \left( \frac{g^\prime}{10^{-2}} \right) \rho_R(T_c), \quad
\left( \frac{g^\prime}{10^{-2}} \right)^{1/2} \gtrsim 4 \times 10^{-4} \left( \frac{v_\Phi}{10^{13}\,\text{GeV}} \right).\label{constset1}
\end{equation}
These two constraints are less stringent than the other constraints discussed later. 

Once the field rolls down to the true vacuum, it oscillates coherently around \( v_{\Phi} \) with angular frequency \( m_\Phi = \sqrt{2\lambda} v_\Phi \), behaving like matter \cite{Datta:2022tab}. {\color{black} Here we have assumed that the field oscillates in the quadratic potential.}  These oscillations persist, and if the scalar field is long-lived, the universe undergoes a prolonged period of matter domination {\color{black} which starts at a temperature $T_{\rm dom}$ and decays at a temperature $T_{\rm dec}$.  }

In our model, the lifetime of \( \Phi \) is determined by its decay into Higgs bosons via the 1-loop coupling \( \lambda_{H\Phi}^{\text{1-loop}} \), which depends on neutrino parameters. Our parameter choices automatically forbid the decay channel \( \Phi \to Z^\prime Z^\prime \), since  

\begin{equation}
M_{Z^\prime} = \sqrt{2} g^\prime v_\Phi > m_\Phi = \sqrt{2 g^\prime} g^\prime v_\Phi.
\end{equation}

Furthermore, we prevent the kinematic decay \( \Phi \to N N \) by ensuring \( M_N \gtrsim m_\Phi \). Consequently, 1-loop decays mediated by off-shell \( N_R \) or \( Z^\prime \) are also forbidden. The induced \( \Phi \Phi Z Z \) coupling via 1-loop kinetic mixing \cite{km1,km2,km3,km4} is significantly weaker than the dominant \( \Phi \Phi h h \) coupling used in our analysis. In this scenario, the most competitive decay channels involve virtual \( Z^\prime \)-mediated 1-loop decays into SM fermions ($f$) and vector bosons ($V$). However, it can be shown that the three-body decay \( \Phi \to f\bar{f}V \) dominates over the two-body decay channels due to the latter being chirality suppressed \cite{Blasi:2020wpy,Han:2017yhy}.

\textbf{Additional Constraints:}  
In addition to the constraints shown in Eq.~\eqref{constset1}, the parameter space of the model is subject to the following stronger constraints:

\begin{enumerate}
    \item \textbf{Bound on the Scale of Leptogenesis:}  
    Since the RHNs acquire mass after the phase transition at \( T = T_c \), the leptogenesis scale \( T_{\rm lepto} \sim M_N \) must satisfy  
    \begin{equation}
        T_{\rm lepto} \sim M_N \lesssim T_c.
    \end{equation}
    This leads to the constraint \( m_\Phi \lesssim M_N \lesssim T_c \), which can be rewritten as\footnote{In numerical computation, we consider a slightly stronger upper bound: \( f_N \lesssim g^{\prime 3/4} \), neglecting terms of \( \mathcal{O}(f_N^4) \) in the effective potential.}  
    \begin{equation}
        \sqrt{2g^\prime}g^\prime \lesssim f_N \lesssim 2\sqrt{g^\prime}. \label{rhmass_b}
    \end{equation}
    
    \item \textbf{Decay Channel Hierarchy:}  
    The decay channel \( \Phi \rightarrow hh \) must dominate (this is crucial to determine the start and the end of the matter domination with neutrino/leptogenesis parameters) over \( \Phi \rightarrow f\bar{f}V \), requiring  
    \begin{equation}
        \Gamma_\Phi^{hh} \gtrsim \Gamma_\Phi^{f\bar{f}V},\label{const2}
    \end{equation}
   where the decay width \( \Gamma_\Phi^{f\bar{f}V} \) is given by~\cite{Datta:2022tab,Blasi:2020wpy}  
    \begin{equation}
        \Gamma_\Phi^{f\bar{f}V} \simeq \lambda g^{\prime 4} \left( \frac{m_\Phi}{10^8~\text{GeV}} \right).
    \end{equation}
    
    \item \textbf{Scalar Field Domination:}  
    The scalar field must dominate the energy density of the universe before it decays, implying  
    \begin{equation}
        T_{\rm dec} < T_{\rm dom},\label{const3}
    \end{equation}
    where \( T_{\rm dom} \) is the temperature at which the scalar field dominates.

     \item \textbf{Upper Bound on \( v_\Phi \):}  
    We impose an upper bound on the vacuum expectation value of \( \Phi \) as  
    \begin{equation}
        v^{\rm max}_\Phi \simeq 5 \times 10^{15}~\text{GeV} \lesssim \Lambda_{\rm GUT}.\label{const4}
           \end{equation}
           
{\color{black}  Upon inclusion of the seesaw Lagrangian, one must account for additional theoretical constraints, namely the seesaw perturbativity condition, given by $\text{Tr}(f_D^\dagger f_D) \lesssim 4\pi$. This condition imposes an upper limit on the symmetry-breaking scale, and consequently on the RHN mass scale. Using neutrino oscillation data to fix the light neutrino masses, this constraint can be recast in terms of physical parameters~\cite{lepgw1,Ghoshal:2022kqp}. It turns out that the resulting perturbative upper bound, $\sim 5 \times 10^{15}~\text{GeV}$, lies below the GUT scale, $\sim 10^{16}~\text{GeV}$. A slightly more conservative condition would be to require that the symmetry-breaking scale remains below the maximum reheating temperature allowed by CMB observations \cite{BICEP:2021xfz,Planck:2018vyg}. Interestingly, both bounds turn out to be numerically comparable.

    \item \textbf{Big Bang Nucleosynthesis (BBN) Constraint:}  
    To be consistent with BBN, the scalar field must decay before BBN begins, ensuring that the Universe has transitioned back to a radiation-dominated phase by temperatures of order $\sim 10~\text{MeV}$. This requirement imposes the condition:
\begin{equation}
    T_{\rm dec} \gtrsim T_{\rm BBN} \sim 10~\text{MeV},
\end{equation}
where $T_{\rm BBN}$ is the temperature at the onset of BBN. }

\end{enumerate}

{\bf Free Parameters:} The model has, in principle, three free parameters: the gauge coupling $g'$, the Yukawa coupling $f_N$, and the vacuum expectation value $v_\Phi$. Instead of treating $v_\Phi$ as a free parameter, we consider $M_N = f_N v_\Phi$ as the independent parameter. However, as explained in Ref.~\cite{Chianese:2024nyw}, the key aspect of this leptogenesis model is to work with a fixed gauge coupling. Specifically, within a given UV (e.g., GUT) framework, the idea is to analyze how the leptogenesis parameter space—particularly the scale $T_{\rm lepto} \sim M_i$—is sensitive to propagating gravitational waves.\\

{\bf Features of the model:}  In the limit \( m_\Phi \gg m_h \), the decay rate of \( \Phi \) is given by~\cite{Gross:2015bea,Enqvist:2016mqj,Croon:2019dfw}
\begin{equation}
    \Gamma_\Phi^{hh} \simeq \frac{({\lambda_{H\Phi}^{\rm 1-loop}})^2 \, v_\Phi^2}{32\pi m_\Phi} 
    \simeq \Gamma_0~\frac{f_N^6}{\lambda} \left(\frac{v_\Phi}{10^{13}~\text{GeV}}\right)^2
    \left(\frac{m_\Phi}{10^8~\text{GeV}}\right)~\ln^2\left(\frac{\Lambda}{\mu}\right)\,,
    \label{newgamma}
\end{equation}
where  \( \Gamma_0 \simeq 1.3\times 10^{-2} \) and the one-loop effective coupling is given by  
\begin{equation}
    \lambda_{H\Phi}^{\rm 1-loop} \sim \frac{f_D^2 f_N^2}{2\pi^2} \ln \left(\frac{\Lambda}{\mu}\right),
\end{equation}
with  
\begin{equation}
    f_D = \sqrt{\frac{m_\nu f_N v_\Phi}{v_h^2}}.
\end{equation}
For numerical estimates, we use \( m_\nu \simeq 0.01~\text{eV} \) and \( v_h = 174~\text{GeV} \).  
The logarithmic factor accounts for the renormalization condition, ensuring that even if it is defined to vanish at a very high scale, such as \( \Lambda \equiv \Lambda_{\rm GUT} \simeq 10^{16}~\text{GeV} \), a nonzero value appears at lower scales.

Considering \( \mathcal{H} \simeq \Gamma_\Phi^{hh} \), the decay temperature \( T_{\rm dec} \) is computed as  
\begin{equation}
    T_{\rm dec} = \widetilde{T}_{\rm dec} \ln\left(\frac{\Lambda}{\mu}\right)\,, \label{eq:tdec}
\end{equation}
where  
\begin{equation}
    \widetilde{T}_{\rm dec} = \left(\frac{90}{\pi^2 g_*}\right)^{1/4} 
    \left[\tilde{M}_{\rm Pl}~\Gamma_0~\frac{f_N^6}{\lambda}
    \left(\frac{v_\Phi}{10^{13}~\text{GeV}}\right)^2
    \left(\frac{m_\Phi}{10^8~\text{GeV}}\right)\right]^{1/2}\,,
\end{equation}
with \( \tilde{M}_{\rm Pl} = 2.4\times 10^{18}~\text{GeV} \) being the reduced Planck mass and \( g_* \simeq 106.75 \) the effective degrees of freedom.

For the physical scale \( \mu \equiv T_{\rm dec} \), the solution to Eq.~\eqref{eq:tdec} is given by  
\begin{equation}
    T_{\rm dec} = \tilde{T}_{\rm dec}~\mathcal{W}\left(\frac{\Lambda}{\widetilde{T}_{\rm dec}}\right)\,, 
    \label{newdec}
\end{equation}
where \( \mathcal{W} ( {\Lambda}/{\widetilde{T}_{\rm dec}} ) \) is the Lambert function, modulating \( T_{\rm dec} \) due to the imposed renormalization condition.

Using Eq.~\eqref{newdec}, the two key quantities for describing gravitational wave spectral features, \( T_{\rm dom} \) and the entropy production factor \( \kappa \) can be obtained as
\begin{equation}
    T_{\rm dom} = \frac{\rho_\Phi(T_c)}{\rho_R(T_c)}T_c\,, \quad 
    \kappa = \frac{T_{\rm dom}}{T_{\rm dec}} \Theta\left[T_{\rm dom}-T_{\rm dec}\right] \,, 
    \label{newentr}
\end{equation}
where the Heaviside function \( \Theta \) ensures that the expression for \( \kappa \) is valid only when \( T_{\rm dom} > T_{\rm dec} \).{ \color{black}Two key quantities defined in Eq.\eqref{newentr}, and used for numerical scanning in this article, have been cross-checked to reproduce results of the Friedmann equations (see, Ref.\cite{Chianese:2024nyw}).}

\begin{figure}[t!]
\hspace{-.8cm}\includegraphics[scale=.28]{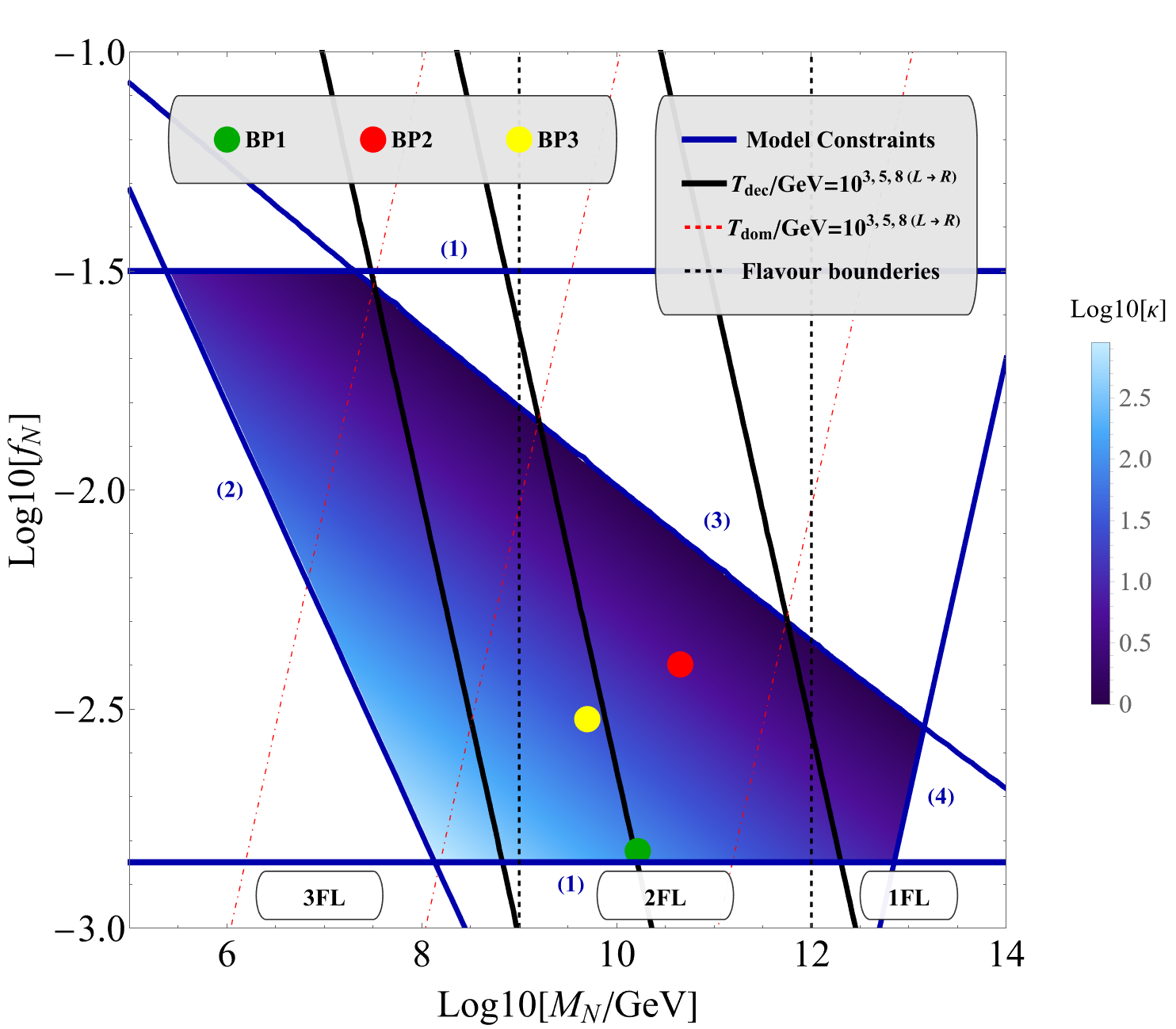}\includegraphics[scale=.28]{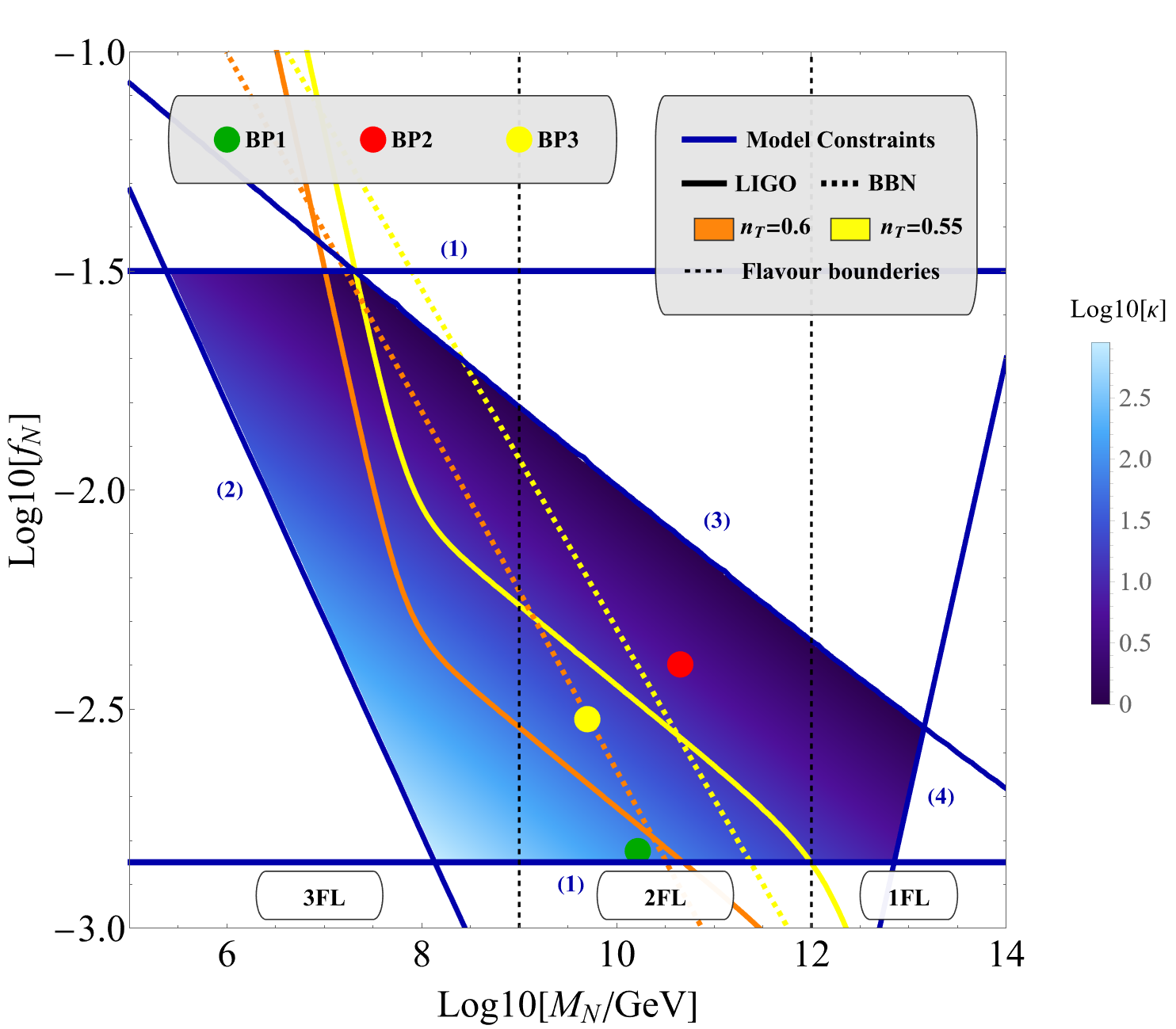}
\caption{Left: the coloured region represents the allowed model parameter space subjected to constraints {\color{black}[Eq.\eqref{rhmass_b}(1), Eq.\eqref{const2} (2), Eq.\eqref{const3} (3), Eq.\eqref{const4} (4)]} described in the main text for $g^\prime=10^{-2}$. The produced entropy $\kappa$ is shown with the blue gradient. The oblique black thick (red dashed thin) lines represent different decay (domination) temperatures. Right: Model parameter space subjected to GW constraints imposed by nonobservation of stochastic gravitational waves by LIGO (solid curves) and BBN bound on effective neutrino species (dashed curves) for different $n_T$ values. The parameter space right to these curves is excluded.}
\label{fig:fig1} 
\end{figure}

We shall work with large values of gauge coupling by fixing $g^\prime=10^{-2}$ motivated by GUT models\footnote{Parameter space discussing smaller/larger values of gauge coupling can be found in Ref.\cite{Chianese:2024nyw}. As an aside, we would also like to mention that this model can also produce thick cosmic strings as discussed in Ref.\cite{Chianese:2024gee}.}. In the left-hand side of Fig.\ref{fig:fig1}, we show the allowed parameter space of the model with relevant constraints described before and the entropy gradient in blue. We show only the constraints (1)-(4), because the scalar field decays much before the BBN; constraint (5).  The BPs for which we performed the Fisher analysis are shown with filled circles and also displayed in Table \ref{t1}.
 
\begin{table}[H]
    \centering 
     \caption{Leptogenesis model parameters }\label{t1}
    \begin{tabular}{c|c|c}
           &~~ $M_N~\left[\rm GeV\right]$~~ & $~~~f_N~~~$\\ \hline
           ~~ $\text{BP1}$ ~~& ~~ $1.65\times 10^{10}$ ~~ &~~ $1.5\times 10^{-3}$\\ 
           ~~ $\text{BP2}$ ~~ & ~~$ 4.5\times 10^{10}$~~ & ~~ $4\times 10^{-3}$\\ 
           ~~ $\text{BP3}$ ~~ & ~~  $5\times 10^{9}$ ~~ & ~~ $3 \times 10^{-3} $
    \end{tabular}
    \end{table}
  {\color{black}
    Since we intend to focus on signals within the two-flavour regime (a brief discussion on the three-flavour regime is presented in Appendix \ref{appn3}), the BPs are well dispersed within this regime. Moreover, they correspond to varying levels of entropy production, which leads to a higher peak-to-dip ratio of GW amplitudes in cases with large entropy production \cite{Chianese:2024nyw}. The BP2 (red) is excluded for spectral indices $n_T > 0.55$ by LIGO and BBN bounds (discussed later). However, we aim to present BPs that are representative of broader trends within the LISA-sensitive parameter space. Therefore, in this case, the LIGO and BBN bounds were considered less strictly. That said, BP2 becomes viable for lower values of $n_T$ (e.g., $n_T < 0.55$), and a similar trend can still be illustrated. }

   The oblique black thick (red dashed thin) lines represent different decay (domination) temperatures, delineating distinct 
flavour regimes of leptogenesis:  
\begin{itemize}
    \item \( T_{\rm dec} \lesssim 10^3 \) GeV for the three-flavour regime (3FL),  
    \item \( 10^5 \text{ GeV} \lesssim T_{\rm dec} \lesssim 10^8 \text{ GeV} \) for the two-flavour regime (2FL),  
    \item \( T_{\rm dec} \gtrsim 10^8 \) GeV for the one-flavour regime (1FL).  
\end{itemize}
{\color{black}As discussed in detail in the next section,} this classification has significant implications for the spectral features of GWs, since the scenario predicts a double-peak GW spectrum \cite{Chianese:2024nyw}, and the low-frequency peak is determined by \( T_{\rm dec} \); \( f_{\rm peak}\propto T_{\rm dec}  \). {\color{black} For a given \( g' \) in a top-down model,  different flavour regimes correspond to distinct ranges of low-frequency peaks. \textbf{Consider also an intermediate case, e.g., \( T_{\rm dec} \sim 10^4 \) GeV}, which lies at the intersection of two regimes. Two points along this line but in different flavour domains correspond to significantly different \( T_{\rm dom} \). In this case, while both regimes may predict the same low-frequency peak, the subsequent dip frequencies and the ratio of GW amplitudes at the peak and at the dip vary \cite{Chianese:2024nyw}.} This occurs because of changes in  \( T_{\rm dom} \), therefore in \( \kappa \). The dip frequency is primarily controlled by \( T_{\rm dom} \), while \( \kappa \) directly impacts the GW amplitude. As a result, in this model, variations in the right-handed neutrino mass—shifting among the flavour regimes—lead to distinct GW signatures. On the right-hand side, we show the model parameter space with the LIGO and   BBN constraints on Stochastic Gravitational-Wave Background (SGWB), as discussed in the next section.

\section{Properties of the leptogenesis model imprinted primordial gravitational waves}\label{s3}
{\color{black}The spectrum of GW energy density relevant for detection purposes is expressed as}
\begin{equation}
\Omega_{\rm GW}(k) = \frac{k}{\rho_c} \frac{{\rm d}\rho_{\rm GW}}{{\rm d}k}\,,
\end{equation}
where $\rho_c = \frac{3 \mathcal{H}_0^2}{8\pi G}$ is the critical density, with $\mathcal{H}_0 \simeq 2.2 \times 10^{-4}~\rm Mpc^{-1}$ being the Hubble constant. The quantity $\Omega_{\rm GW}(k)$ is given by \cite{gwc0,gwc01,gwc1,gwc2,gwc3}
\begin{equation}
\Omega_{\rm GW}(k) = \frac{1}{12\mathcal{H}_0^2} \left(\frac{k}{a_0}\right)^2 T_T^2(\tau_0,k) P_T(k), \quad \tau_0 = 1.4\times 10^4 {\rm ~Mpc}\,.\label{omgGW}
\end{equation}
Here, {\color{black} $P_T(k)$ denotes the power spectrum of the
inflationary tensor perturbation, for which we adopt a power-law parametrization motivated by
CMB observations:}
\begin{equation}
P_T(k) = r A_s(k_*) \left(\frac{k}{k_*}\right)^{n_T}, \label{Pt}
\end{equation}
where $r \lesssim 0.06$~\cite{BICEP2:2018kqh} is the tensor-to-scalar ratio, and $A_s \simeq 2\times 10^{-9}$ is the scalar perturbation amplitude at the pivot scale $k_* = 0.01~{\rm Mpc^{-1}}$. In our analysis, we consider $r = 0.06$ and a constant, blue-tilted tensor spectral index ($n_T > 0$). However, scale dependence may arise due to higher-order corrections~\cite{Kuroyanagi:2011iw,Calcagni:2020tvw}, depending on the inflation model, which we do not discuss here.

The transfer function $T_T^2(\tau_0,k)$ can be computed analytically, matching numerical results with reasonable accuracy~\cite{t1,t2,t3,t4,t5,t6}. When an intermediate matter domination phase is present—an essential aspect of this work involving  RHN mass-dependent early matter domination—it takes the form~\cite{t5,t6}
\begin{equation}
T_T^2(\tau_0,k) = F(k) \,T_1^2(\zeta_{\rm eq}) \, T_2^2(\zeta_{\Phi}) \, T_3^2(\zeta_{\Phi R}) \, T_2^2(\zeta_{R})\,. \label{totT}
\end{equation}
The individual transfer functions are given by
\begin{align}
T_1^2(\zeta) &= 1+1.57\,\zeta+ 3.42 \,\zeta^2\,,\\
T_2^2(\zeta) &= \left(1-0.22\,\zeta^{1.5}+0.65\,\zeta^2 \right)^{-1},\\
T_3^2(\zeta) &= 1+0.59\,\zeta+0.65 \, \zeta^2\,,
\end{align}
where $\zeta_i = k/k_i$. The characteristic modes $k_i$ are defined as
\begin{align}
k_{\rm eq} &= 7.1\times 10^{-2}\Omega_m h^2 {\rm Mpc^{-1}}, \quad \Omega_m = 0.31, \quad h = 0.7,\\
k_{\Phi} &= 1.7\times 10^{14} \left(\frac{g_{*s}(T_{\rm dec})}{106.75}\right)^{1/6} \left(\frac{T_{\rm dec}}{10^7 \rm GeV}\right) {\rm Mpc^{-1}},\label{fp}\\
k_{\Phi R} &= 1.7\times 10^{14} \kappa^{2/3} \left(\frac{g_{*s}(T_{\rm dec})}{106.75}\right)^{1/6} \left(\frac{T_{\rm dec}}{10^7 \rm GeV}\right) {\rm Mpc^{-1}},\label{fd}\\
k_{R} &= 1.7\times 10^{14} \kappa^{-1/3} \left(\frac{g_{*s}(T_{R})}{106.75}\right)^{1/6} \left(\frac{T_{R}}{10^7 \rm GeV}\right) {\rm Mpc^{-1}}.
\end{align}
These modes re-enter the horizon at different cosmological epochs: $T_{\rm eq}$ (standard matter-radiation equality), $T_{\rm dec}$ (scalar field decay), $T_{\rm dom}$ (scalar field domination), and $T_{R}$ (reheating after inflation). In this framework, reheating occurs before the phase transition of $\Phi$, i.e., $T_{R} > T_c$, ensuring that the condition for thermal leptogenesis, $T_{R} > M_N$, is met since $T_c > M_N$ (see Eq.~\eqref{rhmass_b}).

The reheating temperature is thus bounded as $T_c \lesssim T_R \lesssim T_R^{\max}$, where the maximum allowed value is $T_R^{\max} \simeq 10^{15}$ GeV \cite{BICEP:2021xfz,Planck:2018vyg}. In the present analysis choice of $T_R$ does not play a crucial role since it mostly affects the GW spectrum at frequencies much higher than LISA band. We, therefore, work with $T_R=T_c$. { \color{black} For a detailed discussion on the impact of $T_R$ on the parameter space, see Ref.\cite{Chianese:2024nyw} (Fig.4 and Fig.5), also footnote-6 in this article.}

The function $F(k)$ in Eq.~\eqref{totT} is given by
\begin{equation}
F(k) = \Omega_m^2 \left( \frac{g_*(T_{k,\rm in})}{g_{*0}}\right) \left(\frac{g_{*0s}}{g_{*s}(T_{k,\rm in})}\right)^{4/3} \left(\frac{3j_1(k\tau_0)}{k\tau_0}\right)^2, 
\end{equation}
where $T_{k,\rm in}$ is the temperature at which mode $k$ enters the horizon, $j_1(k\tau_0)$ is the spherical Bessel function, and the parameters take values $g_{*0} = 3.36$ and $g_{*0s} = 3.91$. An approximate form of the scale-dependent $g_*$~\cite{gs1,gs2,t6} can be found in Ref.\cite{Chianese:2024nyw}.

The quantity $\Omega_{\rm GW}(k)h^2$ is constrained by two robust bounds on the SGWB, namely from the effective number of relativistic species during BBN~\cite{Peimbert:2016bdg} and LIGO measurements~\cite{LIGOScientific:2016jlg}. The BBN constraint reads
\begin{equation}
\int_{f_{\rm low}}^{f_{\rm high}} {\rm d} f ~f^{-1} \Omega_{\rm GW}(f)h^2 \lesssim 5.6\times 10^{-6} \Delta N_{\rm eff}\,,
\end{equation}
where $\Delta N_{\rm eff} \lesssim 0.2$. The integration lower bound corresponds to the frequency of the mode entering the horizon during BBN, $f_{\rm low} \simeq 10^{-10}$ Hz, while we set the upper limit as $f_{\rm high} \simeq 10^{7}$ Hz, which covers all possible allowed values of $T_R$. For the LIGO constraint, we assume a reference frequency $f_{\rm LIGO} = 25~{\rm Hz}$ and {\color{black}discard GWs with the amplitudes  $\Omega_{\rm GW}h^2>8.33\times 10^{-9}$ at this frequency~\cite{KAGRA:2021kbb}.} Let us stress that the analysis of Ref.\cite{Chianese:2024nyw} and consequently the present article is strictly valid for power-law parameterization of $P_T$, as adopted e.g., in Ref.\cite{bn3n} and also by the PTA collaboration \cite{ng5}. A different parameterization, e.g., see Ref.\cite{Jiang:2023gfe} would yield  different results.
\section{LISA instrumental noise,  foregrounds, and Fisher matrix forecast}\label{s4}
\textbf{LISA Instrumental Noise:} Following Ref.\cite{Caprini:2019pxz}, we adopt a single TDI (Time Delay Interferometry) output noise model (TD1X). However, our results can be easily generalized to a three TD1 channel (either to XYZ or to AET basis) analysis \cite{Flauger:2020qyi}. For example, an optimal channel combination in the AET basis \cite{Flauger:2020qyi} would result in a factor of $\sim 1/\sqrt{2}$ improvement in the uncertainties reported in this paper. The TD1X noise spectrum is given by 
\bea
h^2\Omega_{\rm Inst}^X =h^2\frac{4 \pi^2}{3 \mathcal{H}_0^2}f^3\frac{N_{\rm Inst}^X}{R_X},\label{noiseden}
\eea
where 
\bea
N_{\rm Inst}^X = 16 \sin^2 \left( \frac{2\pi f L}{c} \right)\left\{ P_{\text{oms}}(f, P) + \left[ 3 + \cos \left( \frac{4\pi f L}{c} \right) \right] P_{\text{acc}}(f, A) \right\}\label{noise}
\eea
with $c=3\times 10^8$ m/s, $L=2.5\times 10^6$ km, and we work within the LISA frequency band $f\in \left[3\times 10^{-5}\text{Hz},\,\,0.5\text{Hz}\right]$.
The noise components defined as optical metrology system noise $P_{\text{oms}}(f, P)$  and mass mass acceleration noise $P_{\text{acc}}(f, A)$ are given by
\bea
P_{\text{oms}}(f, P) = P^2 \frac{ \text{pm}^2}{\text{Hz}} 
\left[ 1 + \left( \frac{2 \, \text{mHz}}{f} \right)^4 \right] 
\left( \frac{2\pi f}{c} \right)^2,
\eea

\bea
P_{\text{acc}}(f, A) = A^2 \frac{\text{fm}^2}{\text{s}^4 \text{Hz}} 
\left[ 1 + \left( \frac{0.4 \, \text{mHz}}{f} \right)^2 \right] 
\left[ 1 + \left( \frac{f}{8 \, \text{mHz}} \right)^4 \right] 
\left( \frac{1}{2\pi f} \right)^4
\left( \frac{2\pi f}{c} \right)^2.
\eea
According to the LISA Pathfinder mission \cite{Armano:2018kix} and science requirement document \cite{LISAsc} the noise in the TD1X channel can be established within an uncertainty of $20\%$. Based on this, we adopt the noise parameter values as $P=15$ and $A=3$ \cite{Caprini:2019pxz,Flauger:2020qyi,LISAsc,Robson:2018ifk}. The polarization and sky-averaged response
function (response of a detector to the GW signal) is given by \cite{Caprini:2019pxz,Flauger:2020qyi}
\bea
R_X=16 \sin^2 \left( \frac{2\pi f L}{c} \right) 
\frac{3}{10}\frac{1}{1 + 0.6 \left( \frac{2\pi f L}{c} \right)^2} 
\left( \frac{2\pi f L}{c} \right)^2.\label{rf}
\eea
 Note that the sinusoidal prefactor appearing in Eq.\eqref{noise} and \eqref{rf} cancels in the final expression of $\Omega_{\rm Inst}^X$ in Eq.\eqref{noiseden}.

\textbf{Galactic Foreground:} The Galactic foreground primarily originates from white dwarf binaries within our galaxy \cite{Nissanke:2012eh,Evans:1987qa}. While some of these binaries can be individually resolved and subtracted from the data \cite{Cornish:2017vip}, the unresolved sources form a strong foreground. Following the approach in Refs. \cite{Caprini:2024hue,Blanco-Pillado:2024aca}, we consider the integrated signal across the entire sky dome over the full mission duration and adopt the template from Ref. \cite{Karnesis:2021tsh}:

\begin{equation}
    h^2 \Omega_{\text{Gal}}^{\text{GW}}(f) = \frac{f^{3} }{2}
    \left(\frac{f}{1 \text{ Hz}}\right)^{-7/3} 
    \left(1 + \tanh \left(\frac{f_{\text{knee}} - f}{f_2} \right) \right) 
    e^{-(f / f_1)^{\nu}} h^2 \Omega_{\text{Gal}},
\end{equation}

where

\begin{align}
    \log_{10} (f_1 / \text{Hz}) &= a_1 \log_{10} (T_{\text{obs}} / \text{year}) + b_1, \\
    \log_{10} (f_{\text{knee}} / \text{Hz}) &= a_k \log_{10} (T_{\text{obs}} / \text{year}) + b_k,
\end{align}

and

\begin{align*}
    a_1 &= -0.15, \quad b_1 = -2.72, \\
    a_k &= -0.37, \quad b_k = -2.49, \\
    \nu &= 1.56, \quad f_2 = 6.7 \times 10^{-4} \text{ Hz}, \\
\end{align*}
We consider $\Omega_{\text{Gal}}$ to be a free parameter and use $ \log_{10} (h^2 \Omega_{\text{Gal}}) = -7.84$ \cite{Karnesis:2021tsh} as the fiducial value in the Fisher analysis. Throughout the paper, we shall consider 4 years of effective observation time. 

\textbf{Extragalactic Foreground:} This foreground arises from compact object binaries located outside the Milky Way that LISA cannot resolve individually. The primary contributors are neutron star and white dwarf binaries, along with stellar-origin black hole binaries emitting gravitational waves during their inspiral phase. Due to LISA's limited angular resolution and the relatively uniform distribution of these sources, this foreground is modeled as stationary, Gaussian, and isotropic. It can be distinguished from cosmological signals only through its frequency spectrum. We adopt the following spectral shape for the extragalactic foreground \cite{Regimbau:2011rp,Babak:2023lro,Lehoucq:2023zlt}:

\begin{equation}
    h^2 \Omega_{\text{Ext}}^{\text{GW}}(f) = h^2 \Omega_{\text{Ext}} 
    \left(\frac{f}{1 \text{ mHz}}\right)^{2/3},
\end{equation}
where $\Omega_{\text{Ext}}$ is treated as a free parameter in the Fisher analysis, with a fiducial value given by

\begin{equation}
    \log_{10} (h^2 \Omega_{\text{Ext}}) = -12.38.
\end{equation}
For convenience, we present the noise and foreground fiducial values in Table \ref{t1}. Since our primary focus is on computing uncertainties in the primordial BGW signals imprinted by leptogenesis, we will display error contours exclusively for the primordial signal. Additionally, we will list the fiducial values corresponding to the primordial signal as `signal fiducials' alongside the respective figures.
 
\begin{table}[H]
    \centering 
     \caption{Fiducials: Noise and foreground }\label{t2}
    \begin{tabular}{c|c|c|c}
         $P$\,& $~~A~~$ & $\log_{10} (h^2 \Omega_{\text{Gal}})$ & $~~~\log_{10} (h^2 \Omega_{\text{Ext}})~~~$\\ \hline
         15~~ & ~~ 3 ~~& ~~ -7.84~~ &~~ -12.38\\
        
    \end{tabular}
    \end{table}
\textbf{Signal-to-Noise Ratio:} Like other forecast analyses \cite{Caprini:2019pxz,Flauger:2020qyi,Caprini:2024hue,Blanco-Pillado:2024aca,Gowling:2021gcy,Gowling:2022pzb,Giese:2021dnw,LISACosmologyWorkingGroup:2025vdz,nr1,nr2}, we quantify the signal-to-noise ratio (SNR) for the relevant GW spectra, particularly those with lower uncertainties. In this model, we find that stronger GW spectra—comparable to or exceeding the TD1X noise level in certain frequency ranges—tend to be associated with reduced parameter uncertainties. Unlike the commonly used SNR formula in the weak-signal limit~\cite{Maggiore:1999vm,Schmitz:2020syl,Saikawa:2018rcs,Allen:1997ad,Romano:2016dpx,Kudoh:2005as}, a more accurate estimation, in this case, can be obtained using the general optimal filter function for arbitrary signal strengths, as outlined in \cite{Kudoh:2005as}. The filter function reads  

\bea
\tilde{Q}(f)=\frac{U(f)V(f)^*-U(f)^*W(f)}{|V(f)|^2-W(f)^2}
\eea
that returns SNR as \cite{Kudoh:2005as}
\bea
{\rm SNR}=\sqrt{2T_{\rm obs}}\left[\int_{-\infty}^{+\infty}\frac{df}{2}\frac{|U(f)|^2}{|U(f)|^2+W(f)}\right]^{1/2},\label{snrgen}
\eea
where the frequency-dependent functions are given by 
\bea
U(f) &=& C_{IJ}(f) + \delta_{IJ} N_I(f),\\
V(f) &=& C_{IJ}^2 + \delta_{IJ} N_I \big(N_I + 2C_{II}\big),\\
W(f) &=& C_{II}C_{JJ} + C_{II}N_J + C_{JJ}N_I + N_I N_J
\eea
with 
\bea
C_{IJ}(f; t, t) &=& \int \frac{d\hat{d}}{4\pi} S_h(f) R_{IJ}(f, \hat{d}; t, t),\\
\braket{\tilde{n}^*_I(f) \tilde{n}_J(f')} &=& \frac{1}{2} \delta_{IJ} \delta(f - f') N_I(f)\\
\langle \tilde{h}^*_A(f, \hat{d}) \tilde{h}_{A'}(f', \hat{d}') \rangle &=&
\frac{1}{2} \delta(f - f') \delta^2(\hat{d}, \hat{d}') \frac{\delta_{AA'}}{4\pi} S_h(|f|, \hat{d}).
\eea

The indices $I, J$ represent different detectors, while $R_{IJ}$ is the antenna pattern function, which characterizes the overlap in the detectors' responses to the signal. The terms $\tilde{n}$ and $\tilde{h}$ denote the noise and the GW signal, respectively. The indices $A, A'$ correspond to the polarization states of the GW signal, and $\hat{d}$ represents the direction from which the GW originates. The strain power spectral density $S_h(f)$ is related to $\Omega_{\text{GW}}(f)$ as  
\bea
\Omega_{\text{GW}}(f) = \frac{4\pi}{3} \frac{f^3 S_h(f)}{\mathcal{H}_0^2}.\label{gwstrn}
\eea  

Defining $C_{IJ} \equiv S_h(f) \Gamma_{IJ}$, where $\Gamma_{IJ} \equiv \Gamma_{\rm LISA}$ (which must be identified with $R_X$ in Eq.~\eqref{rf}), the signal-to-noise ratio (SNR) for a single-channel analysis simplifies from Eq.~\eqref{snrgen} to\footnote{This expression for SNR has recently been used for LISA and DECIGO in Ref.\cite{Brzeminski:2022haa}, and slightly modified version for ET in Ref.\cite{Datta:2024bqp}.}
\bea
{\rm SNR}=\sqrt{T_{\rm obs}}\left[ \int_{f_{\rm min}}^{f_{\rm max}} df \left(\frac{\Omega_{\rm GW}(f)^2}{2\Omega_{\rm GW}(f)^2+2\Omega_{\rm GW}(f)\Omega_{\rm noise}(f)+\Omega_{\rm noise}(f)^2}\right)\right]^{1/2},\label{snr}
\eea  
where we define $S_n(f) = \frac{N}{\Gamma_{\rm LISA}}$, which is related to $\Omega_{\rm noise}$ via Eq.~\eqref{gwstrn}. The SNR expression in Eq.~\eqref{snr} reduces to the standard formula widely used in the literature \cite{Maggiore:1999vm,Schmitz:2020syl} in the weak signal limit ($\Omega_{\rm GW} \ll \Omega_{\rm noise}$). In our SNR computation, we assume $\Omega_{\rm noise} = \Omega_{\rm Inst}^X$. However, as discussed in Ref.\cite{Poletti:2021ytu,Racco:2022bwj}, the presence of an astrophysical foreground can potentially reduce SNR. This reduction can arise from specific spectral shapes of GWs and uncertainties in astrophysical parameters. Although we do not account for these finer details in our computation, the preferred primordial signal strength in our discussion is much stronger than the astrophysical one, therefore we do not expect significant modification. Finally we compute the Fisher matrix under Gaussian likelihood approximation as \cite{Caprini:2024hue,Gowling:2021gcy}
\bea
F_{\alpha\beta}=T_{\rm obs}\int\frac{df}{\Omega^{\rm tot}(\theta_{\Vec{P}},\theta_{\Vec{f}},\theta_{\Vec{N}},f)^2}\frac{\partial\Omega^{\rm tot}(\theta_{\Vec{P}},\theta_{\Vec{f}},\theta_{\Vec{N}},f)}{\partial \theta_\alpha}\frac{\partial\Omega^{\rm tot}(\theta_{\Vec{P}},\theta_{\Vec{f}},\theta_{\Vec{N}},f)}{\partial \theta_\beta},\label{fishana}
\eea
where $\Omega^{\rm tot}=\Omega^{\rm Prim}_
{\rm GW}+\Omega^{\rm Gal}_{\rm GW}+\Omega^{\rm Ext}_{\rm GW}+\Omega^{\rm X}_{\rm Inst}$, $\theta_{\Vec{P}}=\left\{ M_N,f_N,r,n_T\right\}$, $\theta_{\Vec{f}}=\left\{ \Omega^{\rm Gal},\Omega^{\rm Ext}\right\}$, and  $\theta_{\Vec{N}}=\left\{ P, A\right\}$. The square root of the diagonal elements of the inverse Fisher matrix (covariance matrix) represents the uncertainties in the concerned parameter. 

\begin{figure}[t!]
\includegraphics[width=\textwidth]{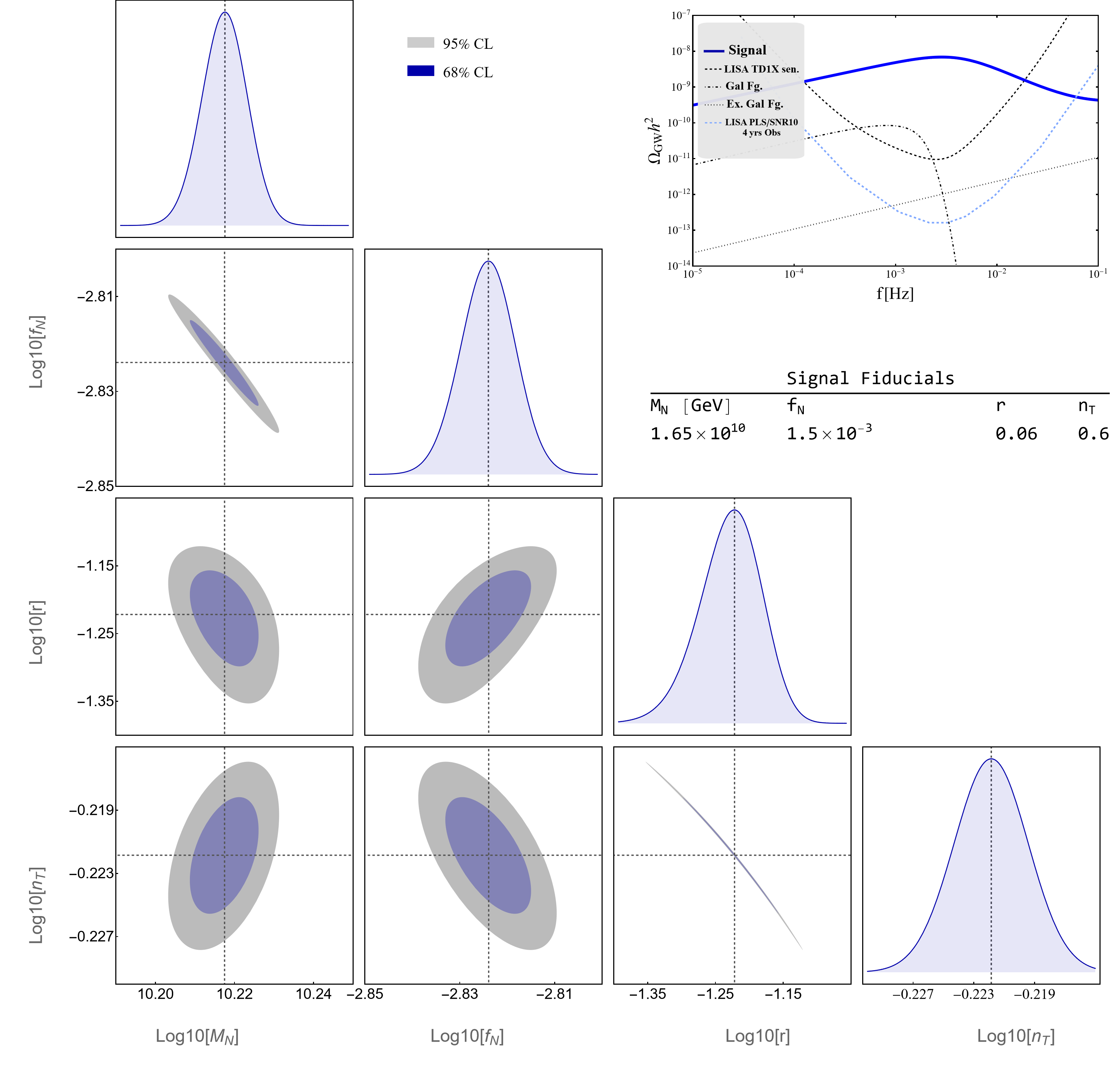}
\caption{(Left) Triangle plots illustrating the uncertainties in BGW signal parameters, assuming perfectly known instrumental noise and astrophysical foregrounds. (Right) The solid blue curve represents the primordial signal corresponding to the fiducial values of the primordial signal parameters (displayed as ``signal fiducials" in the figure) and noise parameters listed in Table \ref{t2}. The dashed black curve denotes the LISA TD1X noise sensitivity, while the dot-dashed black curve represents the galactic foreground. The dotted curve corresponds to extragalactic foregrounds. The dashed blue curve shows the LISA PLS curve for an SNR threshold of 10 over a total observation period of 4 years, computed considering only LISA instrumental noise.}
\label{fig:2} 
\end{figure}
Our final results will be based on studying the full $8\times 8$ covariance matrix, and uncertainties on a given parameter will be computed after marginalizing over the others.  We shall present our result maximizing the primordial signal strength with respect to the tensor-to-scalar ratio: $r\lesssim 0.06$ \cite{BICEP2:2018kqh}. However, as we shall discuss this would be merely a choice, our final projection on the tensor blue-tilt and leptogenesis parameters will remain unchanged. The results obtained from Eq.\eqref{fishana} can also be reproduced with good accuracy accounting for dispersion in the LISA mock data \cite{Caprini:2019pxz,Flauger:2020qyi} corresponding to a fiducial set of values for primordial signal, foreground, and noise parameters, as discussed in the Appendix \ref{appa}. Having set up all the necessary prerequisites, we now discuss the Fisher matrix analysis results in what follows.

\begin{figure}[t!]
\includegraphics[width=\textwidth]{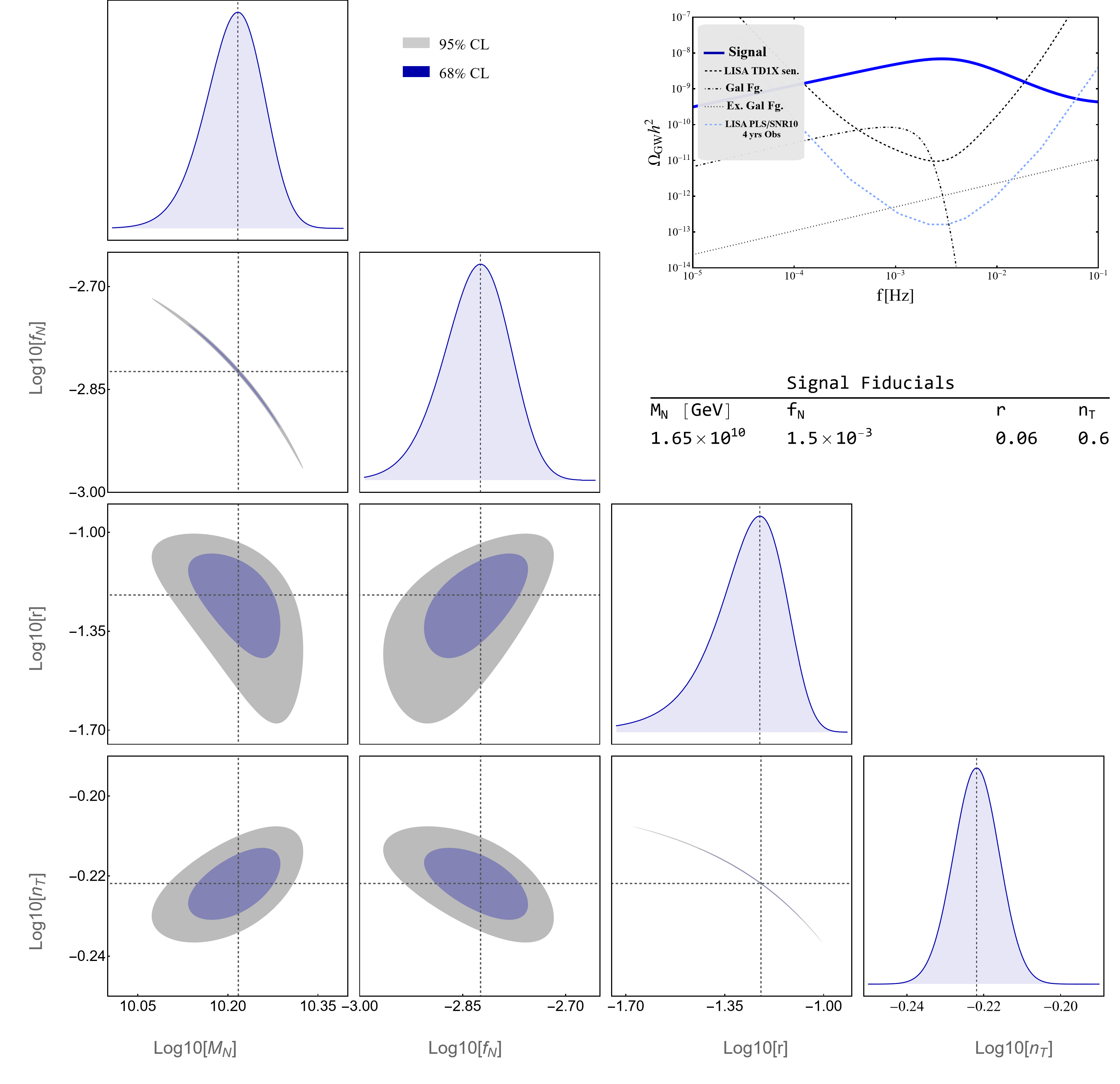}
\caption{(Left) Triangle plots illustrating the uncertainties in BGW signal parameters, accounting for covariance in all eight parameters. (Right) The solid blue curve represents the primordial signal corresponding to the fiducial values of the primordial signal parameters (displayed as ``signal fiducials"  in the figure) and noise parameters listed in Table \ref{t2}. The dashed black curve denotes the LISA TD1X noise sensitivity, while the dot-dashed black curve represents the galactic foreground. The dotted curve corresponds to extragalactic foregrounds. The dashed blue curve shows the LISA PLS curve for an SNR threshold of 10 over a total observation period of 4 years, computed considering only LISA instrumental noise.}
\label{fig:fig3} 
\end{figure}

\textbf{Results:}{ \color{black} We begin with an optimistic scenario }where astrophysical foregrounds and instrumental noise are precisely known. Therefore, we focus on the $4 \times 4$ covariance matrix involving only the primordial signal parameters: $\theta_{\Vec{P}} = \{ M_N, f_N, r, n_T \}$, marginalizing over the remaining three signal parameters to obtain uncertainties for a specific parameter.  

Figure~\ref{fig:2} presents the uncertainty contours in a triangle plot for $n_T = 0.6$ \footnote{\color{black}The choice to focus on values $n_T \lesssim 0.6$ is not set by a fundamental bound, but by observational practicality—this range offers better signal detectability with lower uncertainties in the LISA frequency band. Slightly higher values (e.g., $n_T \sim 0.65$) may still be allowed depending on model parameters, but very large values (e.g., $n_T \gtrsim 0.7$) are typically ruled out by high-frequency constraints, such as LIGO O3, as discussed in Ref.~\cite{Chianese:2024nyw}. This limit depends on the minimal reheating temperature $T_{\rm R} \sim T_C$; for higher $T_{\rm R}$, the bound on $n_T$ tightens, reinforcing $n_T \lesssim 0.6$ as the most suitable choice for both observability and consistency across a range of scenarios.
 }, along with a pair of leptogenesis parameters (BP1) corresponding to a two-flavour regime. The top-right corner of the figure shows the corresponding signal, along with the LISA TD1X noise curve and the Power-law Integrated Sensitivity (PLS) curve (see figure caption for details). This scenario represents a parameter region with minimal uncertainties—less than $10\%$ for $r$ and at the percent level for the other signal parameters.  

\begin{figure}[t!]
\includegraphics[width=\textwidth]{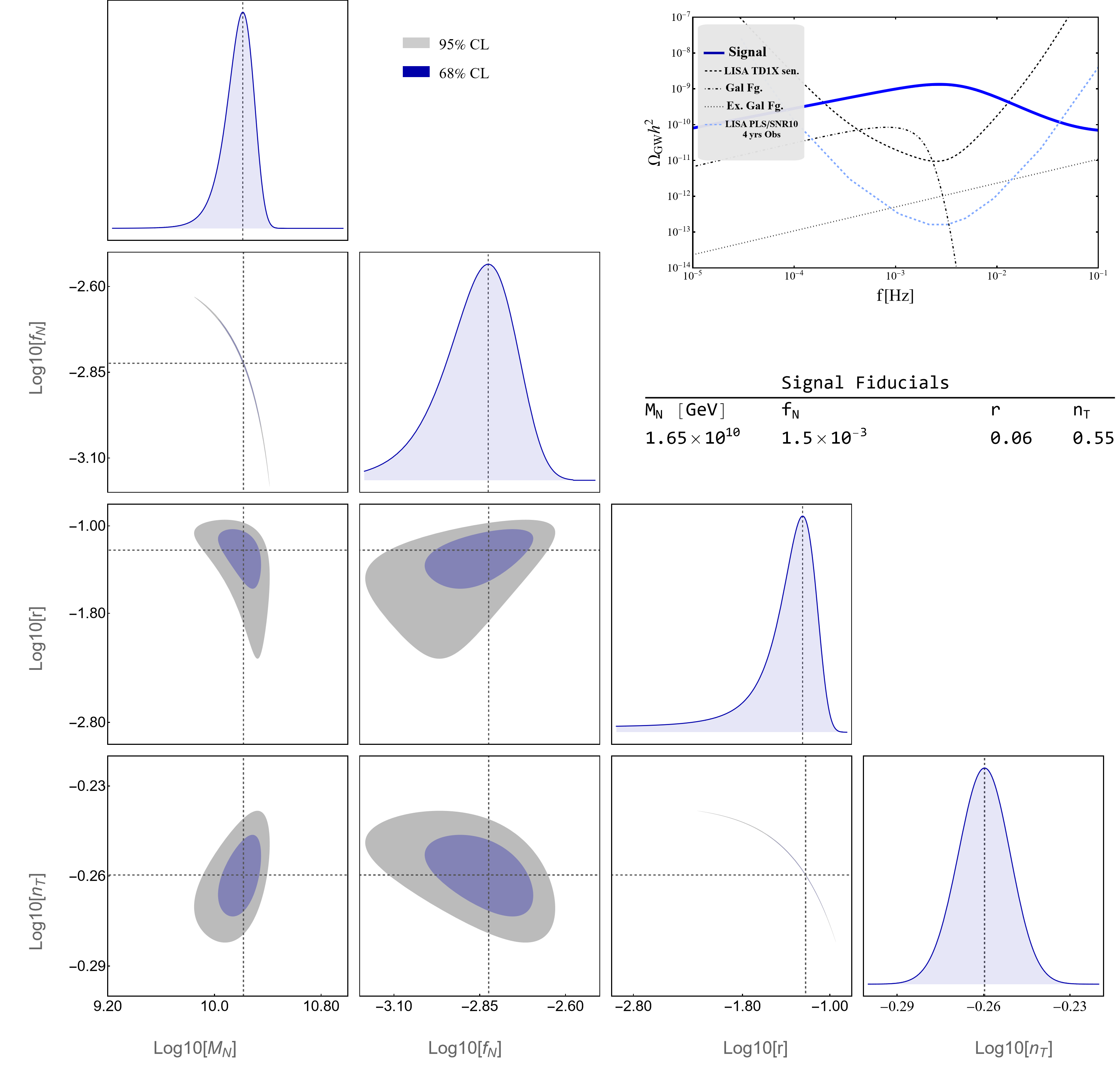}
\caption{(Left) Triangle plots illustrating the uncertainties in BGW signal parameters, accounting for covariance in all eight parameters. (Right) The solid blue curve represents the primordial signal corresponding to the fiducial values of the primordial signal parameters (displayed as ``signal fiducials"  in the figure) and noise parameters listed in Table \ref{t2}. The dashed black curve denotes the LISA TD1X noise sensitivity, while the dot-dashed black curve represents the galactic foreground. The dotted curve corresponds to extragalactic foregrounds. The dashed blue curve shows the LISA PLS curve for an SNR threshold of 10 over a total observation period of 4 years, computed considering only LISA instrumental noise.}
\label{fig:fig4} 
\end{figure}
\begin{figure}[t!]
\includegraphics[width=\textwidth]{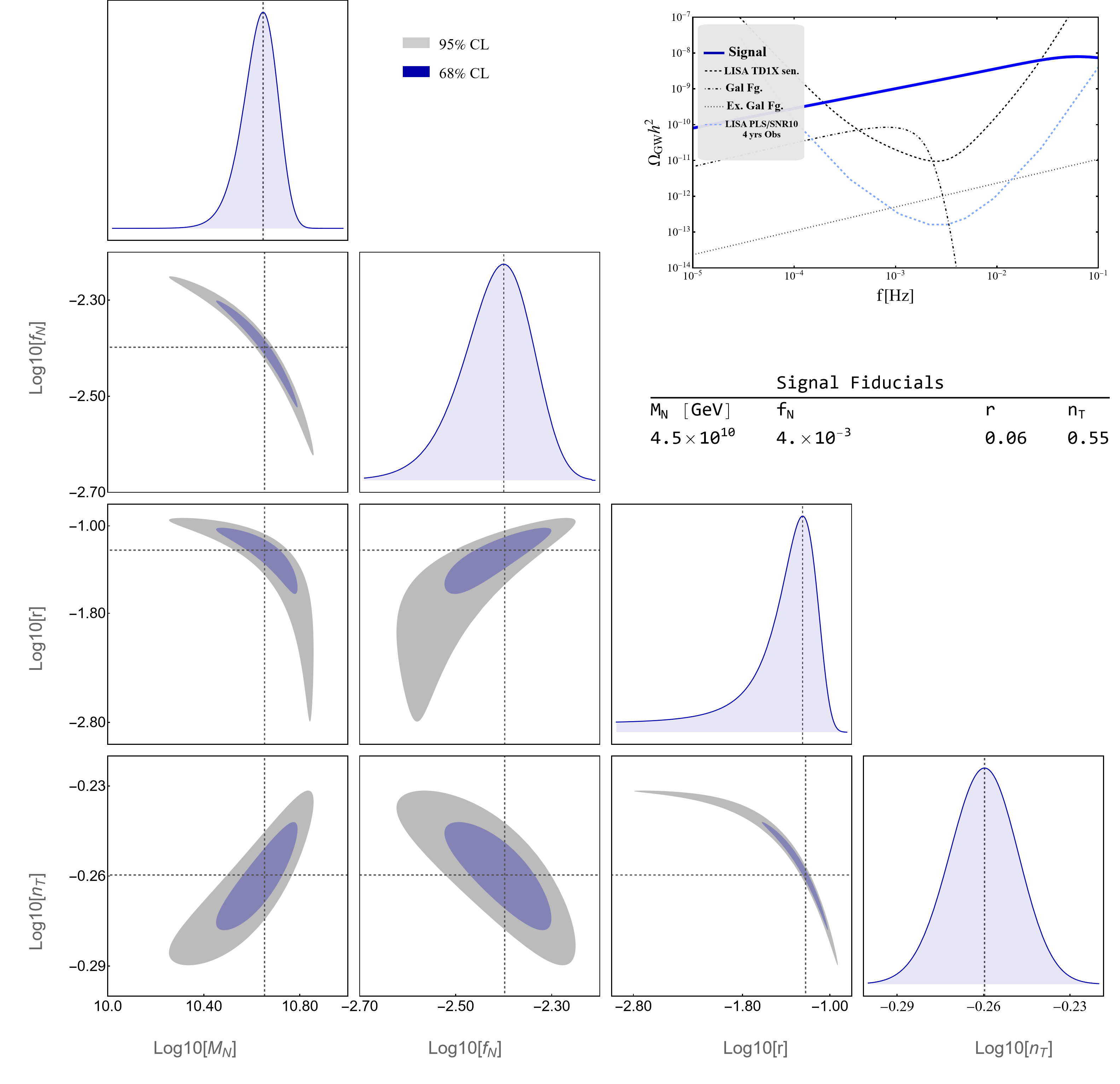}
\caption{(Left) Triangle plots illustrating the uncertainties in BGW signal parameters, accounting for covariance in all eight parameters. (Right) The solid blue curve represents the primordial signal corresponding to the fiducial values of the primordial signal parameters (displayed as ``signal fiducials"  in the figure) and noise parameters listed in Table \ref{t2}. The dashed black curve denotes the LISA TD1X noise sensitivity, while the dot-dashed black curve represents the galactic foreground. The dotted curve corresponds to extragalactic foregrounds. The dashed blue curve shows the LISA PLS curve for an SNR threshold of 10 over a total observation period of 4 years, computed considering only LISA instrumental noise.}
\label{fig:fig5} 
\end{figure}
\begin{figure}[t!]
\includegraphics[width=\textwidth]{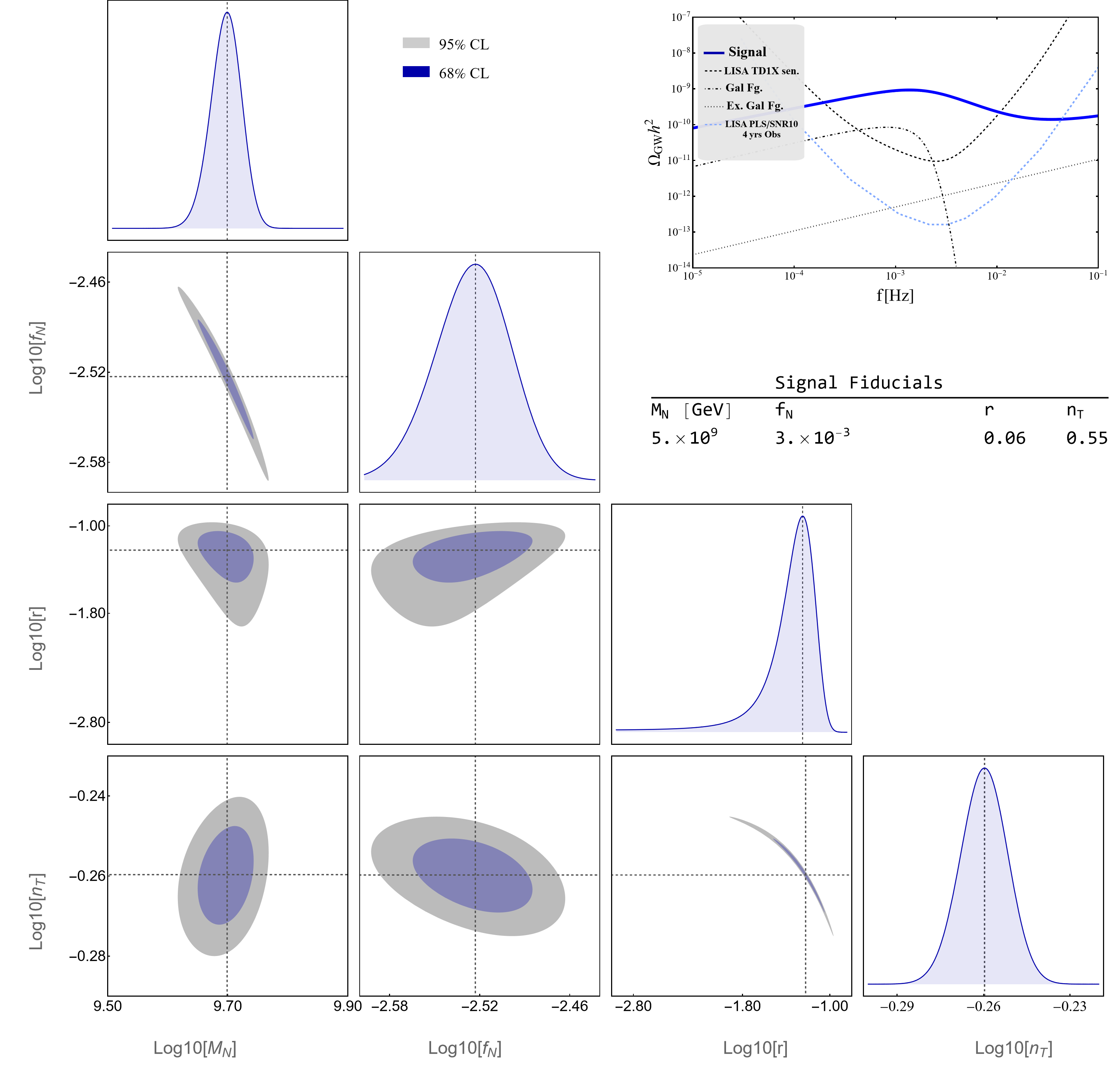}
\caption{(Left) Triangle plots illustrating the uncertainties in BGW signal parameters, accounting for covariance in all eight parameters. (Right) The solid blue curve represents the primordial signal corresponding to the fiducial values of the primordial signal parameters (displayed as ``signal fiducials"  in the figure) and noise parameters listed in Table \ref{t2}. The dashed black curve denotes the LISA TD1X noise sensitivity, while the dot-dashed black curve represents the galactic foreground. The dotted curve corresponds to extragalactic foregrounds. The dashed blue curve shows the LISA PLS curve for an SNR threshold of 10 over a total observation period of 4 years, computed considering only LISA instrumental noise.}
\label{fig:fig6} 
\end{figure}
When astrophysical foregrounds and LISA instrumental noise are precisely known, this model allows for the detection of a GW signal with uncertainties below $50\%$\footnote{Whenever we quote a lower bound on $n_T$, it should be assumed that below that value, at least one signal parameter is associated with more than  $50\%$ uncertainty. In most cases, the larger uncertainty is led by $r$.} for a spectral index of $n_T \sim 0.45$. However, for $n_T \lesssim 0.45$, uncertainties increase, particularly for $M_N$, $f_N$, and $r$, becoming significant for $n_T \sim 0.4$, especially for $r$.

In a more realistic scenario, where uncertainties in the astrophysical parameters $\Omega_{\text{Ext, Gal}}$ are taken into account, the error increases, as illustrated in Fig.~\ref{fig:fig3}. Similar to the previous case, lowering the value of $n_T$ causes the errors in other parameters to grow significantly (see Fig.~\ref{fig:fig4}). Notably, the uncertainty in the parameter $r$ becomes particularly pronounced around $n_T \sim 0.5$. {\color{black} See Appendix \ref{appn2}, for the triangle plots for further lower values of $n_T$.}

One might intuitively expect that increasing $f_N$ could mitigate this issue, given that a larger $f_N$ corresponds to reduced entropy production, leading to less suppression of the gravitational wave signal. However, this is not the case. In this model, $f_N$ spans only an order of magnitude, and the $M_N$–$f_N$ parameter space is constrained by BBN and LIGO bounds (see Fig.~\ref{fig:fig1}). To illustrate this, Fig.~\ref{fig:fig5} and Fig.~\ref{fig:fig6} present two different pairs of leptogenesis model parameters (BP2 and BP3) for $n_T = 0.55$. These figures closely resemble Fig.~\ref{fig:fig4} (in terms of error magnitude), capturing the increase in errors and {\color{black}indicating a poor reconstruction prospect for lower values of $n_T$.  In fact, the value of $r$ could be smaller, which would degrade the signal reconstruction. However, a larger $n_T$ can compensate for this. Therefore, maximizing the signal with respect to $r$ gives the most optimistic lower bounds on $n_T$.}

Overall, a full $8\times8$ covariance analysis in this model suggests that a leptogenesis-imprinted BGW signal can be detected with less uncertainty only if the spectral index $n_T$ is as large as 0.5. Note interestingly that, despite the strong signal strength, the GW signal may still involve significant uncertainties. This closely resembles a recent study on the spectral reconstruction of GWs from a cosmic strings model, which involved a large number of parameters (typically $>4$) describing the fiducial signal \cite{Blanco-Pillado:2024aca}. 

\textbf{Discussion:} 
{\color{black}In our analysis, we consider the simplest scenario where the scalar field rolls at high temperature. In a different region of parameter space, a first-order phase transition could instead lead to a period of vacuum domination with significant e-folding, followed by matter domination (see, e.g., a similar setup in Ref.~\cite{Gonstal:2025qky}). However, because phase transitions remain relatively underexplored in leptogenesis models, particularly within the framework studied here, where the leptogenesis scale can be directly linked to the duration and end of the non-standard expansion phase, we have chosen to introduce the complexities of phase transitions gradually, aiming to maintain a coherent and accessible presentation. A more detailed treatment of this aspect will be provided in an upcoming publication.}

Although the Fisher matrix analysis provides an optimistic estimate, it is known to overestimate uncertainties. A more accurate assessment, potentially yielding lower values of $n_T$, could be obtained using an MCMC simulation. However, after closely inspecting the comparison of Fisher and other methods of posterior sampling in the forecast analyses \cite{Caprini:2019pxz,Flauger:2020qyi,Caprini:2024hue,Blanco-Pillado:2024aca,Gowling:2021gcy,Gowling:2022pzb,Giese:2021dnw,LISACosmologyWorkingGroup:2025vdz},  we do not expect any dramatic changes in the overall results, i.e., we do not expect to recover weak signals described by spectral indices {\color{black}$n_T\lesssim0.4$.  Moreover, the presence of stronger foregrounds from extragalactic white dwarf binaries \cite{Staelens:2023xjn,Hofman:2024xar} would further degrade the prospects for reconstructing weak signals.}

It is also worth noting that while recent PTA data has been fit with $n_T \simeq 1.8$ \cite{ng5}, achieving a tensor blue tilt as large as $n_T = 0.5 - 0.6$ is not straightforward in simple models of inflation and beyond \cite{bgw1,bgw2,bgw3,bgw4,bgw5,bgw6,bgw7,bgw8,bgwnew,Datta:2023xpr}. With this in mind, we think that this model appears to slightly underperform at LISA compared to our previous expectations \cite{Chianese:2024nyw}, where we showed that values as small as $n_T = 0.3$ might allow the detection of a leptogenesis-imprinted BGW signal—although, in that study, we computed only the SNR. 

Nonetheless, this flavoured leptogenesis model besides contributing to the ongoing effort to probe leptogenesis through gravitational waves (see, e.g.,~\cite{Blasi:2020wpy,Chianese:2024gee,lepgw1,lepgw2,lepgw3,lepgw4,Ghoshal:2022kqp,lepgw8,Borah:2022cdx,Chun:2023ezg,Barman:2024ujh,Datta:2024tne,Liu:2025xvm,Borboruah:2024eha}),   serves as a concrete BSM framework capable of producing a broken power-law template signal (see, e.g., Appendix~\ref{appb}) with a strong amplitude detectable by LISA. In other words, if LISA observes a broken power-law signal of the predicted form and strength, this framework would provide a fair explanation. Conversely, weak broken power-law signals at LISA cannot be attributed to the leptogenesis-BGW framework. Additionally, this article presents the first dedicated study on the detectability of gravitational wave signals across all leptogenesis frameworks discussed in the literature so far.

\begin{figure}[t!]
\includegraphics[scale=.29]{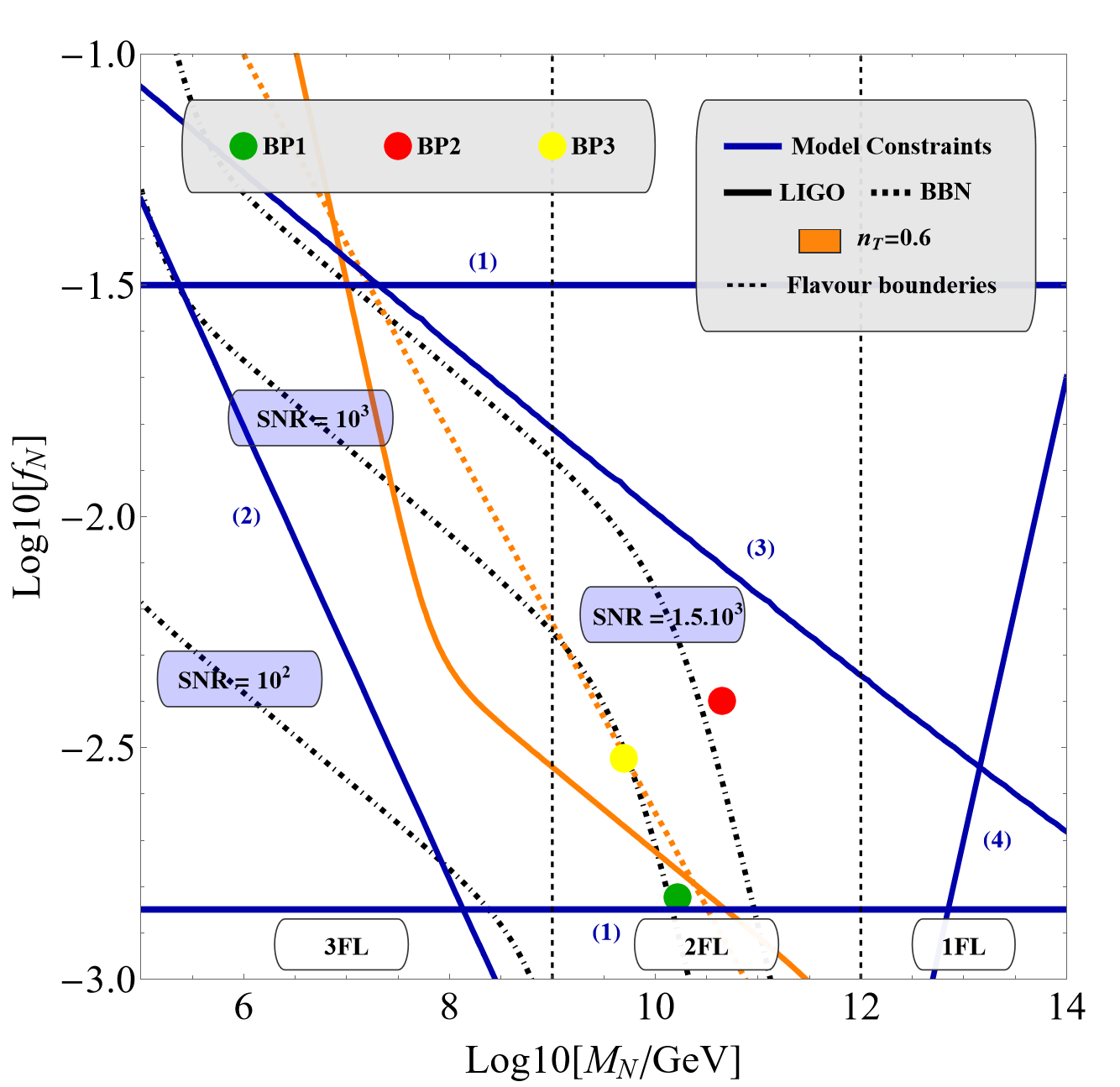}\includegraphics[scale=.29]{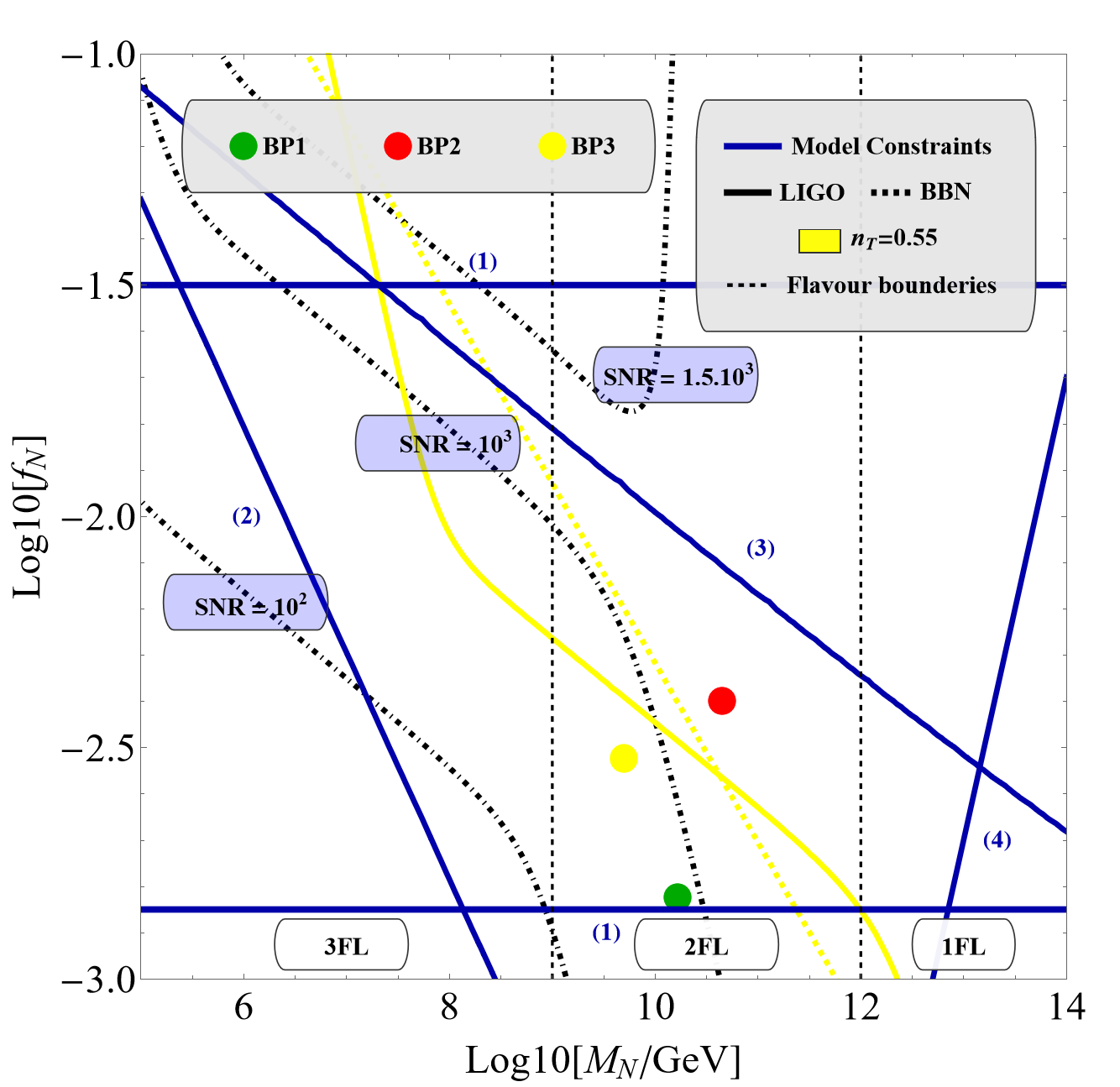}
\caption{With black dashed lines SNR contours are shown on the model parameter space for $n_T=0.6$ (left), $n_T=0.55$ (right). The description of the figure reamins the same as in Fig.\ref{fig:fig1}, barring, this time we remove the entropy gradient from the figures.}
\label{fig:fig7} 
\end{figure}

For completeness, in Fig.~\ref{fig:fig7}, we present the expected SNR on the $M_N$–$f_N$ plane for $n_T = 0.6$ (left) and $n_T = 0.55$ (right). Interestingly, within the allowed parameter space, the expected SNR in this model remains well constrained, with SNR $\in \left[10^2,10^3\right]$.
Finally, we emphasize that although this model predicts a broken—and potentially double-broken—power-law signal within the LISA frequency band, \textbf{such a signal is not a definitive smoking gun for leptogenesis}. This is because the spectral parameters can also be reproduced by other models that feature a period of matter domination with sufficient parameter freedom to achieve $T_{\rm dec}$ and $T_{\rm dom}$ values similar to those in the leptogenesis scenario. Therefore, the validity and robustness of any underlying model can only be assessed statistically.  

In particular, for mock signal reconstruction, given a signal with specific strength and spectral parameters, it is essential to reconstruct the model parameters and compare uncertainties across different models. Similarly, in the case of real data—e.g., if LISA were to observe a broken power-law signal—it would be crucial to perform a Bayesian analysis (or a Fisher analysis for uncertainty contours) to obtain the posterior distributions of the model parameters that best fit the data. By comparing these distributions across different models, one could assess how model-dependent the inferred spectral parameters are. Since the dependence of spectral parameters on model parameters varies among models, different models would likely yield distinct uncertainty distributions.

\section{Conclusion}\label{s5}
The seesaw models based on a Grand Unified Theory (GUT)-embeddable $U(1)_{B-L}$ symmetry exhibit a parameter space that allows for non-standard cosmic expansion. The scalar field responsible for generating the mass of heavy neutrinos can dominate the energy density of the universe in the form of matter. The duration of this matter-dominated phase is determined by the mass of the heavy neutrinos, which typically corresponds to the scale of thermal leptogenesis. Since gravitational waves propagating during a matter-dominated phase retain imprints of that epoch, any gravitational waves generated prior to this phase would carry information about the thermal leptogenesis scale. 

In this work, we consider blue-tilted gravitational waves generated during inflation, which have recently gained attention following the discovery of a stochastic gravitational wave background by pulsar timing arrays. For realistic GUT models with large gauge couplings, our framework predicts a double-peak signal: the first peak appears in the $\mu$Hz region for a three-flavour leptogenesis regime and in the mHz (LISA) region for a two-flavour leptogenesis regime. This article focuses exclusively on the two-flavour leptogenesis scenario, which produces a peak signal (potentially accompanied by a dip) in the LISA frequency band. 

We quantify the uncertainties in the model parameters using the Fisher matrix formalism, incorporating LISA’s instrumental noise along with galactic and extragalactic foregrounds. Our analysis indicates that blue-tilted gravitational waves relating to the leptogenesis model may be detectable by LISA with minimal uncertainties for a tensor spectral index $n_T \gtrsim 0.5$. In this context, such values of $n_T$ correspond to sufficiently strong primordial signals that surpass instrumental noise over a specific frequency range and the foreground strength potentially for the entire frequency band. Thus, if LISA observes a strong amplitude broken power-law signal, this framework would provide a nice interpretation. On the other hand, weak broken power-law signals at LISA cannot be attributed to the discussed leptogenesis-GW framework.

\section*{Acknowledgements}
I would like to thank the referee of \textbf{JCAP 11 (2024) 051}, who anticipated large uncertainties for weak signals and recommended this investigation for this model. I would like to thank  Mauro Pieroni for useful insight regarding parameter reconstruction and Fisher analysis. I would like to thank Marco Chianese, Satyabrata Datta, Gennaro Miele, and  Ninetta Saviano for insightful discussions on building up the model. My work is supported by the research project TAsP (Theoretical Astroparticle Physics) funded by the Istituto Nazionale di Fisica Nucleare (INFN). 
\appendix
{\color{black}
\section{Early matter domination as scale tracers of leptogenesis: A summary for pedestrians}\label{appn1}
Gravitational waves (GWs) provide one of the cleanest and most direct probes of physics operating at extremely high energy scales that lie far beyond the reach of terrestrial colliders. A remarkable feature of GWs is their ability to retain information about the evolution of the early universe, particularly any departure from the standard radiation-dominated equation of state. These deviations manifest in the spectral shape of the stochastic GW background, offering a unique observational window into non-standard thermal histories.

One such non-standard epoch is an early matter-dominated (EMD) era, which can arise naturally in many beyond the Standard Model (BSM) scenarios. The onset, duration, and end of this EMD phase leave characteristic imprints, such as suppression, enhancement, or the appearance of spectral features like double peaks, on the GW spectrum. If these features can be quantitatively connected to physical parameters of specific BSM models, such as symmetry-breaking scales or mass thresholds, then GWs can serve as precision probes of high-scale physics, independent of collider-based experiments.
\begin{figure}[H]
    \centering
    \includegraphics[width=1\linewidth]{Figures/cartoonsigw.pdf}
   \caption{\color{black}
Illustration of how an EMD epoch can encode information about the leptogenesis scale, subsequently imprinted on the GW spectrum. The filled purple circles indicate key cosmological events: reheating after inflation at \( T_R \), scalar field oscillation and RHN mass generation at \( T_c \), leptogenesis at \( T_{\rm lepto} \), onset of EMD at \( T_{\rm dom} \), scalar decay at \( T_{\rm dec} \), and the onset of BBN at \( T_{\rm BBN} \sim 10\,\mathrm{MeV} \). The EMD phase may end before or after the electroweak transition (\( T_{\rm EW} \)). GW signals from sources like inflation, cosmic strings, or scalar-induced perturbations inherit distinct spectral features tied to this thermal history and the leptogenesis scale.
}

    \label{figApp1}
\end{figure}

In this work, we focus on a minimal class of leptogenesis models embeddable in Grand Unified Theories, in which the parameters describing the EMD epoch, particularly its duration and the end, are directly tied to the leptogenesis scale, i.e., the mass of the right-handed neutrinos (RHNs) responsible for generating the baryon asymmetry. As we demonstrate, this connection opens up a novel avenue for probing high-scale leptogenesis using next-generation GW detectors.

\vspace{2mm}

Figure~\ref{figApp1} illustrates three probable pathways through which GWs can encode information about the leptogenesis scale:
\begin{enumerate}
    \item \textbf{Blue-tilted primordial GWs (BGWs)}: A primordial GW background with a blue-tilted spectrum (e.g., arising from certain inflationary models) is distorted by the presence of an EMD phase. The resulting spectrum features a suppression followed by a bump, with the position of the feature tied to the end and duration of EMD—hence, to the leptogenesis scale. This scenario has been formulated in Ref.\cite{Chianese:2024nyw}, and is the focus of the present paper as well.
    
    \item \textbf{Cosmic strings}: In the considered leptogenesis model, the spontaneous breaking of a \( U(1) \) symmetry at high scales can lead to the formation of cosmic strings. The GW background from string networks inherits spectral features associated with the underlying scale of symmetry breaking—again, potentially tied to the leptogenesis scale. This has been discussed in Ref.\cite{Chianese:2024gee}.
    
    \item \textbf{Scalar-induced GWs from early structure formation}: Recently, it has been proposed that early metastable structure (halo) formation, seeded during an EMD epoch, can enhance scalar perturbations that subsequently source second-order GWs \cite{Fernandez:2023ddy}. The frequency and amplitude of the resulting scalar-induced GWs are sensitive to the duration and end of EMD, and thereby to the leptogenesis scale \cite{Chianese:2025mll}.
\end{enumerate}

\vspace{2mm}
 \begin{figure}[t!]
\includegraphics[width=\textwidth]{Figures/p6t88_Finale.png}
\caption{\color{black}(Left) Triangle plots illustrating the uncertainties in BGW signal parameters, accounting for covariance in all eight parameters. The contours, which are not filled with colors, represent a case where the corresponding $1\sigma$ uncertainty exceeds 50$\%$. In this case, it is $r$.  (Right) The solid blue curve represents the primordial signal corresponding to the fiducial values of the primordial signal parameters (displayed as ``signal fiducials"  in the figure) and noise parameters listed in Table \ref{t2}. The dashed black curve denotes the LISA TD1X noise sensitivity, while the dot-dashed black curve represents the galactic foreground. The dotted curve corresponds to extragalactic foregrounds. The dashed blue curve shows the LISA PLS curve for an SNR threshold of 10 over a total observation period of 4 years, computed considering only LISA instrumental noise. }
\label{fig:14} 
\end{figure}

Together, these mechanisms demonstrate the rich potential of gravitational wave cosmology in testing and constraining high-scale baryogenesis scenarios, particularly those like high-scale leptogenesis, which are otherwise difficult to access experimentally.
  }
 
{\color{black}
\section{Prospect of signal reconstruction for three flavour leptogenesis scenario: Signals with a dip in the LISA band}\label{appn3}
In the main text, we discussed the prospects for signal reconstruction in the two-flavour leptogenesis scenario, where the corresponding gravitational wave signals typically feature peaks within the LISA frequency band. In contrast, signals arising from the three-flavour scenario can exhibit dips in the LISA band. After examining a representative set of benchmark points, well dispersed across the parameter space (as shown in Fig.~\ref{fig:fig1}; benchmark points for the three-flavour case are not displayed), we find that such signals generally show poor reconstruction prospects within the LISA frequency range. This is primarily because, in the three-flavour scenario, the BGW signal is more strongly contaminated by noise and astrophysical foregrounds compared to the two-flavour case.

In Fig.~\ref{fig:14}, we present one illustrative benchmark case with $M_N = 4 \times 10^8$~GeV (other parameter values are indicated in the figure as signal fiducials). Although the spectral index is relatively large, $n_T \sim 0.6$, the uncertainty in the $r$ parameter exceeds 50\%. This results in significantly reduced prospects for detecting signals from the three-flavour leptogenesis scenario within the LISA band.
}

\section{Fisher forecast with LISA mock data}\label{appa}
To generate a realization of a simulated signal,  at each value of frequency we generate the quantities
\begin{equation}
S_i = \left|{\frac{G_{i1} \left(0, \sqrt{h^2 \Omega_{\text{GW}}(f_i)} \right) + i G_{i2} \left(0, \sqrt{h^2 \Omega_{\text{GW}}(f_i)} \right)}{\sqrt{2}}} \right|^2\label{randsig}
\end{equation}
\begin{equation}
N_i = \left|{\frac{G_{i3} \left(0, \sqrt{h^2 \Omega_{\text{N}}(f_i)} \right) + i G_{i4} \left(0, \sqrt{h^2 \Omega_{\text{N}}(f_i)} \right)}{\sqrt{2}}} \right|^2\label{randnoise}
\end{equation}
within the LISA  frequency band with a frequency resolution $\Delta f=10^{-6}$. Here $S_i$ includes the primordial signal as well as the foregrounds.
In Eq.\eqref{randsig} and Eq.\eqref{randnoise}, \( G_{i1} (\mu, \sigma) \), \( G_{i2} (\mu, \sigma) \), \( G_{i3} (\mu, \sigma) \), and \( G_{i4} (\mu, \sigma) \) representing the real and imaginary parts of the Fourier coefficients of the signal and noise are 4 real numbers randomly drawn from a Gaussian distribution with mean \( \mu \) and variance \( \sigma^2 \). The signal and noise power values are then combined to generate the data, assuming that the noise is uncorrelated with the signal:
\begin{equation}
D_i = S_i + N_i
\end{equation}
\begin{figure}[t!]
\includegraphics[width=\textwidth]{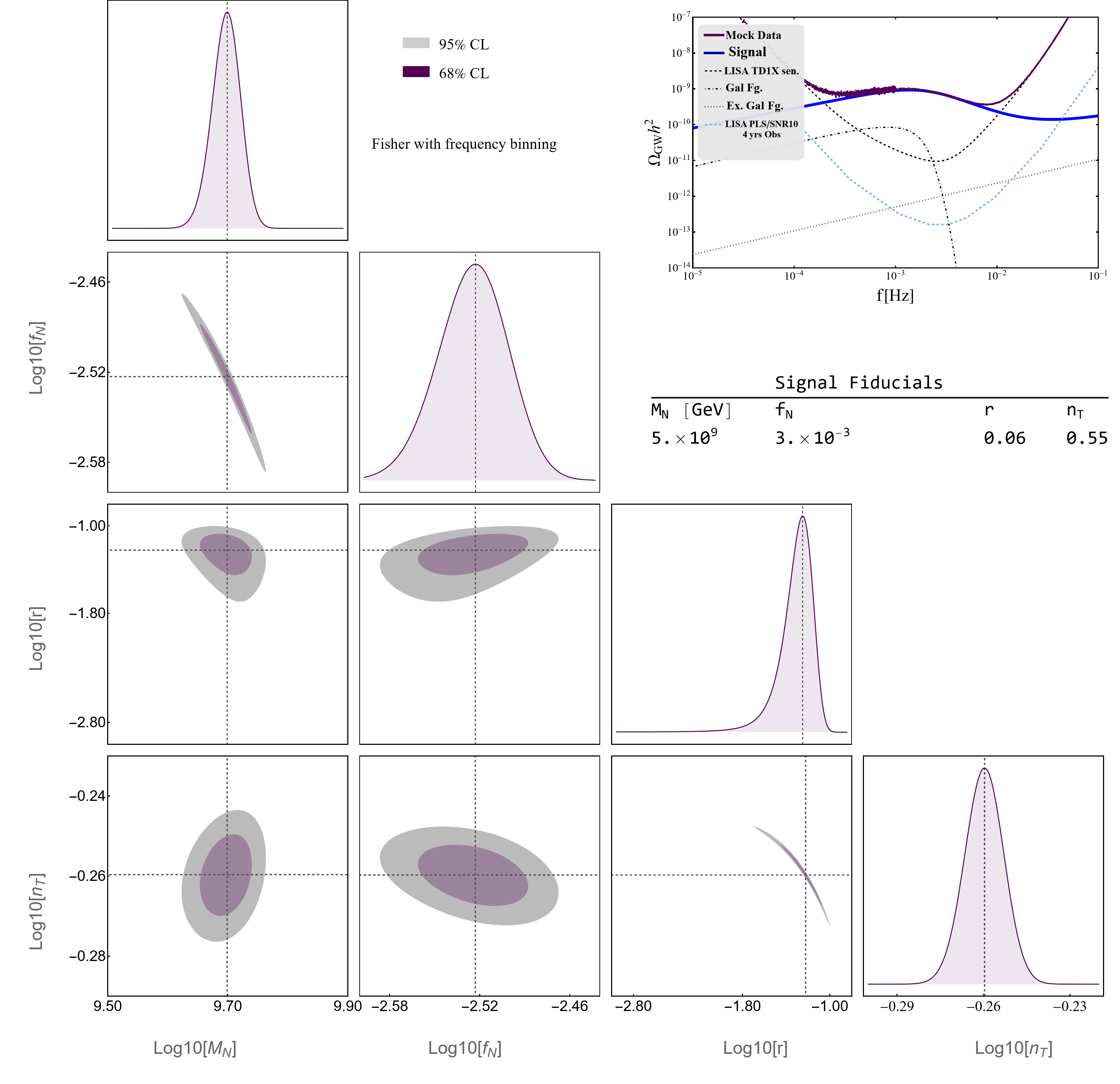}
\caption{(Left) Triangle plots illustrating the uncertainties in BGW signal parameters, accounting for covariance in all eight parameters. (Right) The solid blue curve represents the primordial signal corresponding to the fiducial values of the primordial signal parameters (displayed as ``signal fiducials"  in the figure) and noise parameters listed in Table \ref{t2}. The dashed black curve denotes the LISA TD1X noise sensitivity, while the dot-dashed black curve represents the galactic foreground. The dotted curve corresponds to extragalactic foregrounds. The dashed blue curve shows the LISA PLS curve for an SNR threshold of 10 over a total observation period of 4 years, computed considering only LISA instrumental noise. The scattered points in purple represents LISA mock data generated with Eq.\eqref{randsig} and \eqref{randnoise}.}
\label{fig:binfish} 
\end{figure}
 For each frequency \( f_i \), we generate $N_C=126$ values, \( \{D_{i1}, \dots, D_{i126} \} \), and then compute their mean \( \bar{D}_i \) and standard deviation \( \sigma_i \). The quantity $N_C=126$ represents the fact that within 4 years of observation time, each frequency will be measured  126 times. 
 Given the mock data for  fiducial signal and noise parameter, the signal reconstruction and associated uncertainties can be computed with the likelihood function
 \bea
 \mathcal{L}\left(D|\Omega^{\rm tot}(\theta_{\Vec{S}},\theta_{\Vec{N}})\right) =\prod_{i} \left(\frac{N_C}{2\pi \sigma_i^2}\right)^{\frac{1}{2}}\text{exp}\left[-\frac{N_C}{2}\frac{\left(D_i-\Omega^{\rm tot}(\theta_{\Vec{S}},\theta_{\Vec{N}},f_i)\right)^2}{\sigma_i^2}\right]. \label{lik}
 \eea
We assume that the fiducial signal corresponds to the maximum likelihood of Eq.\eqref{lik}. In the realm of signal reconstruction, this study is equivalent to assessing the quality of the signals to be injected and discarding the model parameter space at the injection level, when associated with large uncertainties (e.g., this study is equivalent to generating the red contours on the triangle plot in Fig.3 in Ref.\cite{Caprini:2024hue}, not the blue one). The Fisher information matrix in the Gaussian approximation then reads 

\bea
F_{\alpha\beta}=\sum_i^b\frac{N_C}{\sigma_i^2}\frac{\partial\Omega^{\rm tot}(\theta_{\Vec{S}},\theta_{\Vec{N}},f_i)}{\partial \theta_\alpha}\frac{\partial\Omega^{\rm tot}(\theta_{\Vec{S}},\theta_{\Vec{N}},f_i)}{\partial \theta_\beta}.\label{binfish}
\eea
The upper limit of the sum $b$ is in principle $b=(0.5-3\times10^{-5})/10^{-6}\simeq 5\times 10^5$. However, in a realistic situation, performing the sum in Eq.\eqref{binfish} and the case of an MCMC simulation with the likelihood function defined in Eq.\eqref{lik} is computationally expensive for such a large number of frequencies. This can be avoided by introducing a statistically equivalent technique by coarse-graining the data within the frequency band $f\in \left[10^{-3}\text{Hz},0.5\text{Hz}\right]$ as discussed, e.g.,  in Refs.\cite{Caprini:2019pxz,Flauger:2020qyi}. Therefore the sum in Eq.\eqref{binfish} can be performed in two segments. In the first segment, $i$ runs from 1 to $b=(10^{-3}-3\times10^{-5})/10^{-6}=970$, and in the second segment, i.e., within the frequency band $f\in \left[10^{-3}\text{Hz},0.5\text{Hz}\right]$, $i$ runs over the coarse-grained frequencies as described below.

We sample the data in the second segment by dividing $\left[10^{-3}\text{Hz},0.5\text{Hz}\right]$ into $k=500$ logarithmic bins containing $n_k$ number of frequencies in each bin. We then  define a new data set $\left\{(D^{(cg)}_k,\sigma^{(cg)}_k),f_k^{(cg)}\right\}$ corresponding to the $k$th bin. The coarse-grained data are obtained as 
\bea
D^{(cg)}_k&=&\sum_j^{n_k} \frac{\sigma_j^{-2}}{\sum_j^{n_k} \sigma_j^{-2}} D_j,\\
\sigma^{(cg)}_k&=& \sqrt{ \frac{1}{\sum_j^{n_k} \sigma_j^{-2}}},\\
f^{(cg)}_k&=&\sum_j^{n_k} \frac{\sigma_j^{-2}}{\sum_j^{n_k} \sigma_j^{-2}} f_j.\\
\eea
The Fisher matrices in the two segments are then given by
\bea
F^{(1)}_{\alpha\beta}=\sum_{i=1}^{970}\frac{N_C}{\sigma_i^2}\frac{\partial\Omega^{\rm tot}(\theta_{\Vec{S}},\theta_{\Vec{N}},f_i)}{\partial \theta_\alpha}\frac{\partial\Omega^{\rm tot}(\theta_{\Vec{S}},\theta_{\Vec{N}},f_i)}{\partial \theta_\beta}\Theta\left[10^{-3}-f_i\right],\label{binfish2}
\eea
and
\bea
F^{(2)}_{\alpha\beta}=\sum_{k=1}^{500}\frac{N_C}{\sigma^{2,(cg)}_k}\frac{\partial\Omega^{\rm tot}(\theta_{\Vec{S}},\theta_{\Vec{N}},f^{(cg)}_k)}{\partial \theta_\alpha}\frac{\partial\Omega^{\rm tot}(\theta_{\Vec{S}},\theta_{\Vec{N}},f^{(cg)}_k)}{\partial \theta_\beta}\Theta\left[f_i-10^{-3}\right],\label{binfish3}
\eea
respectively. The final Fisher matrix is then obtained as 
\bea
F_{\alpha\beta}=F^{(1)}_{\alpha\beta}+F^{(2)}_{\alpha\beta}.
\eea
Note that the choice of the maximum value of \( k \) should align with the desired accuracy of the parameters. Selecting a value greater than 500 can enhance computational accuracy. However, in either case, it is crucial to precisely track the values of \( n_k \) within each frequency bin.

Figure \ref{fig:binfish} presents the error contours using the triangle plot, corresponding to the same fiducial values as in Figure \ref{fig:fig6}. This demonstrates that the Fisher results derived from Eq.\eqref{fishana} can be reproduced using the frequency binning method, with discrepancies of up to 10\% in a few parameters. These discrepancies primarily arise due to coarse-graining and mock data dispersion. The accuracy of the two methods can be improved by either maintaining the original frequency resolution (\(\Delta f = 10^{-6}\) Hz) across the entire frequency range or increasing the number of bins in the second segment during coarse-graining.

The description of the figure in the top right corner remains consistent with previous figures, except for the addition of purple points representing mock data. These points correspond to a frequency resolution of \(\Delta f = 10^{-6}\) Hz up to \( f = 10^{-3} \) Hz, with coarse-graining applied at higher frequencies.

{\color{black}
We conclude this section with the following remark. Although we stated that the uncertainties in the noise parameters are around 20\% (as reported in LISA mission documentation), our analysis is based on a Fisher matrix formalism. In this approach, uncertainties are computed by evaluating numerical derivatives of the likelihood function around a chosen fiducial point and integrating over the LISA frequency band (see, e.g.,  Eq.~\eqref{fishana}/Eq.~\eqref{binfish}). We use the best-fit (preferred) values of the noise and foreground parameters as the fiducial values in our calculation.

\vspace{2mm}

It is important to emphasize that the Fisher matrix approach is a \textbf{frequentist} framework and, unlike \textbf{Bayesian} methods, it does not formally introduce priors. Instead, one constructs the Fisher matrix using expressions such as Eq.~\eqref{fishana}, and obtains parameter uncertainties from the inverse of this matrix, evaluated around a benchmark point. Nevertheless, one can construct ``prior" like quantities on a given parameter from the diagonal elements of the Fisher matrix. They will be equivalent to providing a Gaussian prior around the target values in an MCMC-based Bayesian analysis. Within the LISA band, such uncertainties for all noise and foreground parameters remain below 10\%.
}

{\color{black}
\section{Large uncertainties for lower spectral indices}\label{appn2}
Figures~\ref{fig:12} and~\ref{fig:13} extend the analysis presented in Figures~\ref{fig:2},~\ref{fig:fig3}, and~\ref{fig:fig4}, focusing on cases with lower spectral indices. They demonstrate that the error contours continue to broaden as \( n_T \) decreases, with the most pronounced increase appearing in the uncertainty of the \( r \) parameter. Figures~\ref{fig:12} and~\ref{fig:13} correspond to scenarios with known foregrounds and foregrounds with uncertainties, respectively. For the benchmark case considered, as discussed in the main text, the uncertainty in \( r \) exceeds 50\% for \( n_T \sim 0.45 \) and \( n_T \sim 0.5 \) in the respective cases.

}
\begin{figure}[t!]
\includegraphics[width=\textwidth]{Figures/p444_Finale.png}
\caption{\color{black}(Left) Triangle plots illustrating the uncertainties in BGW signal parameters, assuming perfectly
known instrumental noise and astrophysical foreground. (Right) The solid blue curve represents the primordial signal corresponding to the fiducial values of the primordial signal parameters (displayed as ``signal fiducials"  in the figure) and noise parameters listed in Table \ref{t2}. The dashed black curve denotes the LISA TD1X noise sensitivity, while the dot-dashed black curve represents the galactic foreground. The dotted curve corresponds to extragalactic foregrounds. The dashed blue curve shows the LISA PLS curve for an SNR threshold of 10 over a total observation period of 4 years, computed considering only LISA instrumental noise.}
\label{fig:12} 
\end{figure}
\begin{figure}[t!]
\includegraphics[width=\textwidth]{Figures/p5n88_Finale.png}
\caption{\color{black}(Left) Triangle plots illustrating the uncertainties in BGW signal parameters, accounting for covariance in all eight parameters. (Right) The solid blue curve represents the primordial signal corresponding to the fiducial values of the primordial signal parameters (displayed as ``signal fiducials"  in the figure) and noise parameters listed in Table \ref{t2}. The dashed black curve denotes the LISA TD1X noise sensitivity, while the dot-dashed black curve represents the galactic foreground. The dotted curve corresponds to extragalactic foregrounds. The dashed blue curve shows the LISA PLS curve for an SNR threshold of 10 over a total observation period of 4 years, computed considering only LISA instrumental noise.}
\label{fig:13} 
\end{figure}
\section{A double broken power-law leptogenesis-BGW template}\label{appb}
\begin{figure}[t!]
\includegraphics[scale=.48]{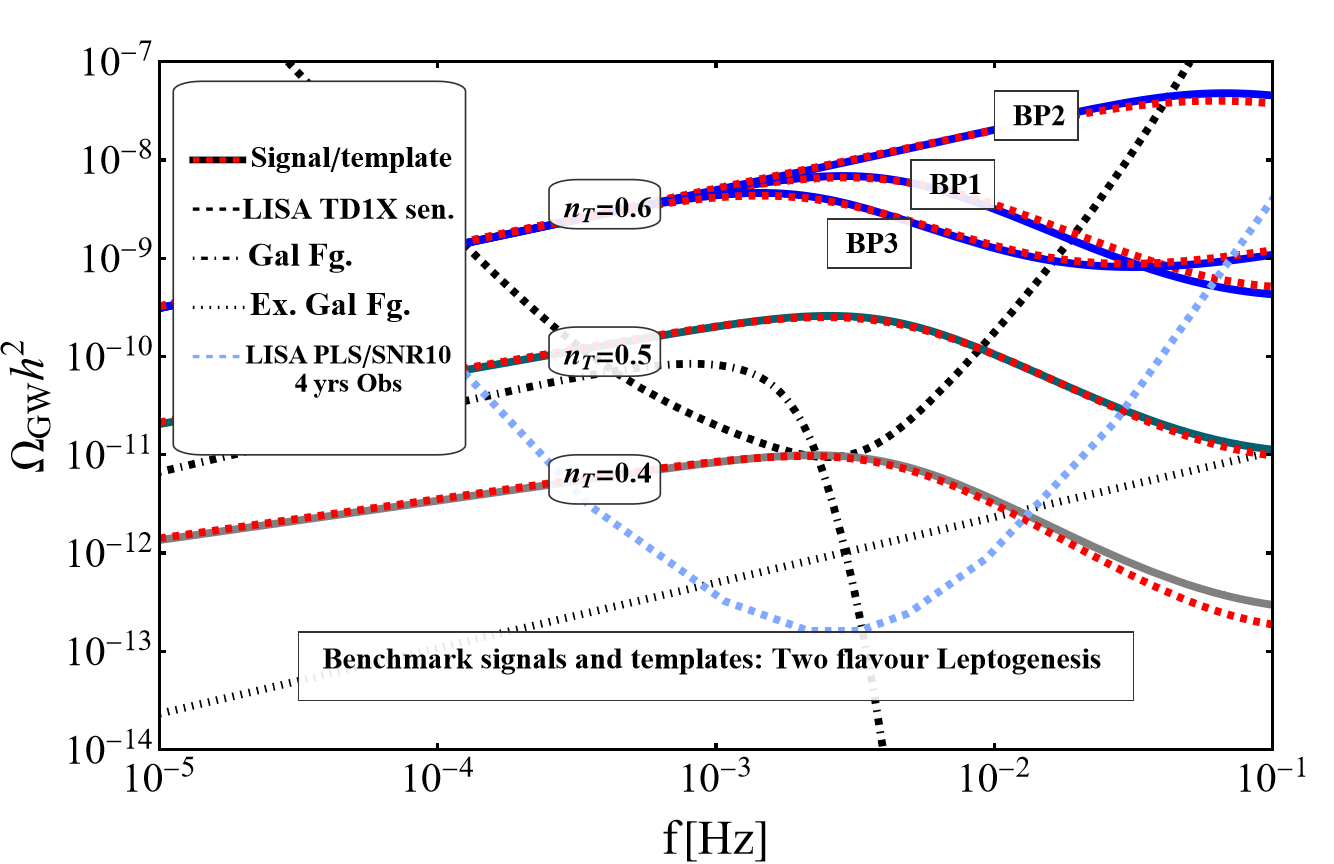}
\caption{ Template GW spectra (dashed) and the model GW spectra (solid). For $n_T=0.6$ all BPs, otherwise BP1 has been shown. The rest of the description of this figure resembles Fig.\ref{fig:fig}.}
\label{fig:temp} 
\end{figure}
Here we present a leptogenesis-BGW template that might be useful for template-based reconstruction studies with the LISA simulator \cite{Caprini:2019pxz,Flauger:2020qyi}. We find that the GW spectrum discussed in Sec.\ref{s3} can be well reproduced within the LISA band with the following template: 
\bea
\Omega_{\text{GW}}^{\text{temp}}( f)h^2 = 
\mathcal{A} \,
r  \left(\frac{f}{f_{\text{CMB}} }\right)^{n_T} 
 \left(\frac{1}{1 + 0.5 \left(\frac{f}{f_{\text{peak}}}\right)^{-(n_T - 2.5)}}\right) 
 \left(1 + 0.6 \left(\frac{f}{f_{\text{dip}}}\right)^{-(n_T - 2.4)}\right),
\eea
where $\mathcal{A}\simeq 5\times 10^{-16}$, $f_{\rm CMB}\simeq (2\pi)^{-1} 10^{-16}$ Hz. {\color{black} The two spectral break frequencies $f_{\text{peak}}$ and $f_{\text{dip}}$ computed from Eq.\eqref{fp} and Eq.\eqref{fd} are given by
\bea
f_{\rm peak} &=& 2.7\times 10^{-3} \left(\frac{g_{*s}(T_{\rm dec})}{106.75}\right)^{1/6} \left(\frac{T_{\rm dec}}{10^5 \rm GeV}\right) {\rm Hz},\label{fbrk1}\\
f_{\rm dip} &=& 5.8\times 10^{-2}  \left(\frac{g_{*s}(T_{\rm dec})}{106.75}\right)^{1/6} \left(\frac{T_{\rm dec}}{10^5 \rm GeV}\right)^{1/3} \left(\frac{T_{\rm dom}}{10^7 \rm GeV}\right)^{2/3}  {\rm Hz}.\label{fbrk2}
\eea
}
Note that the Leptogenesis model information is incorporated into the template spectra through these spectral break frequencies. In Fig.\ref{fig:temp}, we show the model GW spectra (solid) and the template spectra (red dashed) $n_T=0.6$ (all BPs), $n_T=0.5$ (BP1) and  $n_T=0.4$ (BP1). 
\bibliography{bibliography}
\end{document}